\documentclass[journal=jacsat,manuscript=article]{achemso}
\usepackage[utf8]{inputenc}

\usepackage{achemso} 

\setkeys{acs}{maxauthors = 0}
\SectionNumbersOn
\setkeys{acs}{maxauthors=0} 
\usepackage[version=3]{mhchem} 
\usepackage{caption}
\makeatletter
\renewcommand*{\acs@author@fnsymbol}[1]{\ifnum#1=\z@ *\else\number#1\fi}
\renewcommand*\acs@contact@details{%
  { \sffamily *E-mail: \acs@email@list }%
  \acs@number@list
}
\makeatother

\usepackage{xcolor}

\newcommand{\appropto}{\mathrel{\vcenter{
  \offinterlineskip\halign{\hfil$##$\cr
    \propto\cr\noalign{\kern2pt}\sim\cr\noalign{\kern-2pt}}}}}

\author{Pieter J. van Essen}
\affiliation[1]
{Advanced Research Center for Nanolithography, Science Park 106, Amsterdam, 1098 XG, The Netherlands}
\author{Torben Ham}
\affiliation[1]
{Advanced Research Center for Nanolithography, Science Park 106, Amsterdam, 1098 XG, The Netherlands}
\author{Zhonghui Nie}
\affiliation[1]
{Advanced Research Center for Nanolithography, Science Park 106, Amsterdam, 1098 XG, The Netherlands}
\alsoaffiliation[2]{Imaging Physics, Faculty of Applied Sciences, Technische Universiteit Delft, Building 22, Lorentzweg 1, 2628 CJ Delft, The Netherlands}
\author{Peter M. Kraus}
\affiliation[1]
{Advanced Research Center for Nanolithography, Science Park 106, Amsterdam, 1098 XG, The Netherlands}
\alsoaffiliation[3]
{Department of Physics and Astronomy, and LaserLaB, Vrije Universiteit, De Boelelaan 1105, Amsterdam, 1081 HV, The Netherlands}
\email{kraus@arcnl.nl}

\title[Transient High-Harmonic Generation from Solids]
  {Transient High-Harmonic Generation from Solids: From Spectroscopy and Control to Programmable Emission}

\begin{document}

\noindent \textit{All authors contributed equally to this work.}\\[2ex]
High-harmonic generation (HHG) from solids has emerged as a powerful probe of electronic structure and ultrafast dynamics in condensed matter systems. In this chapter, we review recent advances in solid-state HHG spectroscopy and control, with a focus on semiconductors and correlated materials. We discuss the theoretical foundations of HHG and summarize experimental approaches for extracting information from harmonic emission. HHG has been shown to be sensitive to a wide range of microscopic properties, including band structure, transition dipole moments, Berry phases, crystal symmetry, carrier populations, dephasing processes, and lattice dynamics. This sensitivity enables access to electronic and structural dynamics on femtosecond and attosecond timescales and has established HHG as a versatile spectroscopic tool.

At the same time, the strong dependence of HHG on many coupled material and excitation parameters makes the emitted harmonics intrinsically difficult to associate with a single microscopic quantity. Rather than representing a limitation alone, this sensitivity also provides numerous pathways for control. We review how photoexcitation, multicolor driving fields, waveform engineering, and transient modification of material properties can be used to enhance, suppress, and shape harmonic emission. These developments position solid-state HHG not only as a spectroscopic technique, but also as a platform for programmable nonlinear optics with applications ranging from compact extreme-ultraviolet sources to super-resolution microscopy and ultrafast photonic devices.




\section{Introduction}

High-harmonic generation (HHG) is a hallmark process of strong-field physics, first observed in gases in the late 1980s \cite{mcpherson1987,ferray88a}. It provided the basis for attosecond science, where electron dynamics are both generated and probed on sub-cycle timescales. The extension of HHG to solids occurred comparatively late. Theoretical work already anticipated that strong-field-driven currents in crystals could give rise to high-harmonic emission \cite{golde2008}. Experimental realization followed in 2011, when Ghimire \emph{et al.} demonstrated HHG from a bulk semiconductor under intense mid-infrared fields \cite{ghimire2011}. The delay between the discovery of gas and solid HHG is largely explained by the requirement of high-field, long-wavelength driving sources that allow strong-field interaction while avoiding damage.

However, subsequent work quickly established the generality of the process. HHG was observed across a wide range of materials, including semiconductors and dielectrics, and under increasingly relaxed driving conditions. In particular, it was shown that even relatively short wavelengths (e.g. the standard 800~nm wavelength of a Ti:Sa laser) and standard femtosecond laser systems can drive HHG in solids \cite{luu2015,Luu2018e}. This demonstrated that solid-state HHG is not restricted to a narrow parameter regime, but rather reflects a generic nonlinear response of driven electronic systems.

Microscopically, HHG in solids is commonly described in terms of strong-field-driven electronic dynamics in the crystal potential. Rather than treating the nonlinear response only through an effective susceptibility, microscopic models resolve how the driving field creates, accelerates, and recombines electronic excitations, and how the resulting time-dependent currents and polarizations emit high harmonics \cite{golde2008,Vampa2014a,Ghimire2019}. Depending on the material, driving conditions, and harmonic energy, different microscopic contributions may dominate, including nonlinear intraband motion, non-resonant and resonant interband polarization, and multiband transitions. Propagation of the driving and harmonic fields through the sample can reshape the emission strongly, but it is a macroscopic effect and not a microscopic mechanism of harmonic generation, and we treat it separately below. This microscopic perspective provides a direct link between the emitted harmonics and the underlying electronic structure, while also making clear that HHG is not governed by a single universal mechanism.

This description also connects naturally to conventional nonlinear optics. In the perturbative regime, harmonic generation is usually expressed phenomenologically as a Taylor expansion of the polarization in powers of the driving field \cite{boyd2008nonlinear}. The microscopic current-based picture can be viewed as a more explicit description of the same light-matter interaction, particularly valuable when the field strength becomes large enough that the response is no longer well captured by low-order susceptibilities alone. The microscopic picture therefore does not replace perturbative nonlinear optics, but extends its interpretation by resolving the electronic motion, band structure, scattering, and excitation pathways that underlie the nonlinear polarization \cite{aversa1995}. In this sense, one of the main achievements of solid-state HHG since 2011 has been to renew the connection between nonlinear optical observables and microscopic electronic dynamics in solids.

The sensitivity of HHG to electronic dynamics makes it a powerful probe of transient processes in solids. The emitted spectrum and its dependence on driving conditions reflect the instantaneous band structure and the evolution of electronic populations and coherences. This capability builds on concepts established in gas-phase HHG, where the emitted harmonics encode information about molecular structure and orbitals \cite{itatani2004,Vozzi2011a}, as well as their dynamics when probing chemical reactions \cite{Woerner2010a,kraus2013a} or coupled electronic-nuclear dynamics.\cite{Baykusheva2014} In solids, the analogous sensitivity extends to band dispersion, symmetry, and many-body effects.\cite{Vampa2015a,You2017,juergens2024linking,Hohenleutner2015}

Several key observables enable this connection. The harmonic spectrum reflects the accessible energy differences in the band structure, while the emission yield and cutoff depend on the details of carrier acceleration. More generally, the nonlinear response is sensitive to modifications of the electronic structure induced by excitation, such as band renormalization, carrier injection, and dephasing. This provides access to ultrafast dynamics on femtosecond to attosecond timescales.

Beyond conventional band structure \cite{Vampa2015a}, HHG has been shown to probe more subtle properties of solids. These include geometric and topological aspects of the electronic wavefunctions, such as Berry phases \cite{Luu2018a,Silva2019}, as well as symmetry-dependent responses that manifest in polarization and selection rules\cite{You2017,juergens2024linking}. In addition, coupling to lattice degrees of freedom enables access to phonon-driven dynamics and structural changes.\cite{Zhang2024}

Taken together, these aspects establish HHG as a versatile probe of electronic and structural dynamics in solids.

At the same time, high-harmonic generation in solids is increasingly emerging not only as a probe, but also as a controllable and engineerable emission process. Because HHG depends sensitively on both the driving waveform and the microscopic electronic dynamics, multicolor excitation, waveform synthesis, and transient material modification can be used to actively tailor the emitted harmonics. Recent work has demonstrated strong enhancement and suppression of harmonic emission, phase control, polarization shaping, and pathway-selective modulation across a wide range of solids. These developments suggest that solid-state HHG can be understood in terms of programmable nonlinear emission pathways, where the interference of microscopic excitation channels determines the emitted spectrum and wavefront. Such control concepts are particularly attractive because they enable all-optical switching on femtosecond and potentially sub-cycle timescales, while simultaneously opening routes toward compact short-wavelength light sources, petahertz optoelectronics, and label-free super-resolution microscopy\cite{Murzyn2024,murzyn2026}.

The same sensitivity that enables both spectroscopy and control also poses an important challenge. Harmonic spectra, phases, and polarization states can all respond strongly to changes in band structure, carrier populations, dephasing, symmetry, propagation, and the driving waveform. As a consequence, HHG is generally not a selective probe of a single material parameter. Different microscopic changes may produce similar signatures in the harmonic yield, and meaningful interpretation therefore often requires multidimensional measurements, including amplitude, phase, polarization, angular dependence, and pump--probe dynamics, together with careful comparison to microscopic theory.

At the same time, this strong parameter correlation can also become an advantage. While it complicates the direct inversion of HHG observables into unique microscopic quantities, it implies that many distinct physical mechanisms can efficiently modulate harmonic emission. Processes such as carrier excitation, enhanced dephasing, transient band-structure renormalization, symmetry breaking, and multicolor waveform control may all lead to strong enhancement or suppression of HHG. This convergence makes high-harmonic emission intrinsically susceptible to external control and therefore particularly attractive for programmable nonlinear optics. In this sense, the richness and non-universality of solid-state HHG may be some complication for spectroscopy, but it certainly is a key opportunity for engineering ultrafast switchable and controllable emission processes.

\section{Theory of Solid High-Harmonic Generation and Transient High-Harmonic Generation}
\subsection{Solid High-Harmonic Generation}
\begin{figure}[H]
    \centering
    \includegraphics[width=\linewidth]{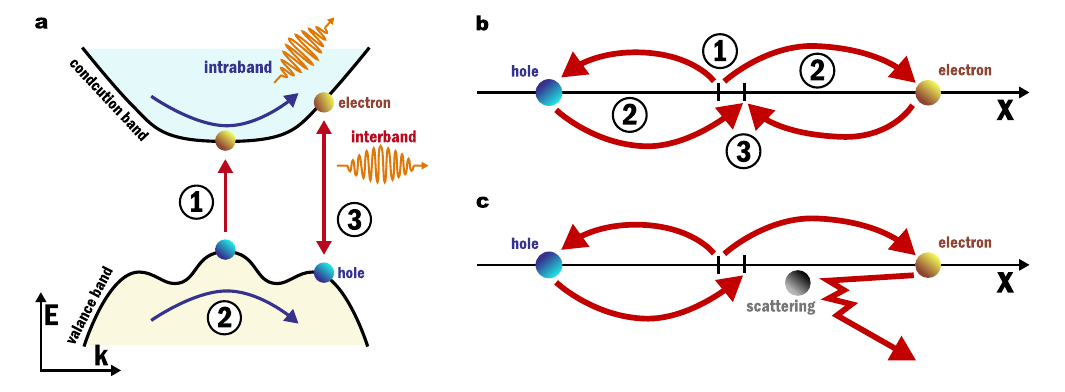}
    \caption{Schematic representations of the three-step model in a solid in reciprocal-space (a) and real-space (b). The three steps are, respectively, excitation, acceleration, and recombination. (c) illustrates schematically how scattering causes dephasing between the electron-hole pair. Adapted from ref \citenum{essen2024HHGcontrol}. Copyright 2024 American Chemical Society under CC BY 4.0, https://creativecommons.org/licenses/by/4.0/.}
    \label{fig:TSM}
\end{figure}

Similar to gas HHG, the basic principles of solid HHG can be explained in terms of a three-step model \cite{Corkum1993,Lewenstein1994,Vampa2017,yue2022}, where HHG is described in terms of subsequent excitation, acceleration, and recombination. The three-step model is schematically shown in Fig. \ref{fig:TSM}a,b in respectively reciprocal and real-space. In the first step, the strong field deforms the potential landscape of the material, enabling excitation of electrons from the valence to the conduction band, which leaves holes behind in the valence band. After excitation, the electron and the hole remain Bloch quasiparticles of the periodic lattice, with a dispersion set by their respective bands and subject to screening and residual Coulomb interaction. They are therefore not free particles in a continuum, as the electron is in the atomic case after strong-field ionization, but are quasi-free carriers whose motion is constrained by the band structure. Within the driving field they are accelerated through reciprocal space, which is the second step. This acceleration of carriers within their bands constitutes the intraband current. That current exists for any band dispersion, including a parabolic one, and its nonlinearity has several origins: non-parabolicity of the bands is the one most often invoked, because the carrier velocity then ceases to follow the vector potential linearly, but a nonlinear intraband current is equally produced by the time-dependent injection of carriers and by their redistribution in reciprocal space during the pulse. As the direction of the driving field flips, the carrier motion can bring the electron and the hole back together, and their recombination contributes to the interband current. Recombination is one pathway that generates an interband polarization, but it is not the only one, as discussed after Eq. \ref{eq:currents}. The intraband and interband currents together form the harmonic emission. Although a simplified picture of the HHG process in solids, this semi-classical interpretation provides a useful framework from which to interpret the more complex simulations.\\
Describing all the many-body interactions in a solid during the HHG process is far outside the scope of current computational capabilities; thus, HHG simulations work with several rigorous approximations. Commonly, the material potential is considered to be static $V(\mathbf{r},t) = V(\mathbf{r})$, the carrier-carrier interactions are neglected, and the problem is treated non-relativistically. Additionally, the dipole approximation $F(\mathbf{r},t) = F(t)$ is employed as the spatial extent of the carrier motion is much smaller than the driving wavelengths. In the length gauge where the vector potential $\mathbf{A}(t) = 0$ and electric scalar potential $\Phi = -\mathbf{r}\cdot \mathbf{F}(t)$, the Hamiltonian in atomic units is given by:
\begin{equation}
    \hat{H}(\mathbf{\hat{r}},t) = \frac{1}{2}\mathbf{\hat{p}}^2+ V(\mathbf{r}) + \mathbf{r}\cdot \mathbf{F}(t).
    \label{eq:Hamil}
\end{equation}
Here $\mathbf{\hat{p}}$ is the momentum operator. Contrary to the perturbative case, the field term is not small compared to the static potential. A variety of implementations exist to solve the equation of motion described by this Hamiltonian, which can be roughly separated into three types: single-particle time-dependent Schr\"{o}dinger equation (TDSE) models \cite{Wu2015,Li2019reciprocal}, density matrix models which include semiconductor Bloch equation (SBE) models \cite{yue2022,Lindberg1998,Silva2019,silva2019PRB}, and time-dependent density functional theory (TDDFT) models \cite{TancogneDejean2017impact,Bauer2018,Jensen2021}. Both single-particle TDSE and density matrix simulations require a description of the material's band structure, energies, and eigenstates as input. For detailed simulations of specific materials, this is obtained using density functional theory (DFT), although a significant number of works make use of simplified model band structures to minimize the computational complexity.\\

In order to illustrate the dynamics underlying the HHG process, we here explicitly show the SBEs that express the equation of motion in terms of the density matrix $\rho$. The diagonal elements $\rho_{m,m}$ describe the population of the bands, while the off-diagonal elements $\rho_{m,n}$ describe the coherence between bands. A detailed derivation of the SBEs in various bases can be found in Ref. \cite{yue2022}. We note that, besides the commonly used reciprocal basis, such as the Bloch basis, recent works have illustrated the benefits of instead using a real-space Wannier basis \cite{molinero2025,silva2019PRB,Silva2019}. Here we consider the SBEs in the length gauge and in the Bloch basis,
\begin{equation}
    \begin{split}
        i \frac{\partial}{\partial t}\rho_{mn}(\mathbf{k},t) =&
    \Big(\epsilon_{m}(\mathbf{k})-\epsilon_{n}(\mathbf{k})\Big)\rho_{mn}(\mathbf{k},t)\\
    &
    - \mathbf{F}(t) \cdot \sum_l\Big(\mathbf{d}_{ml}(\mathbf{k})\rho_{ln}(\mathbf{k},t) - \mathbf{d}_{ln}(\mathbf{k})\rho_{ml}(\mathbf{k},t)\Big) \\
    &
    + i \mathbf{F}(t)\cdot \nabla_\mathbf{k}\rho_{mn}(\mathbf{k},t) -i\frac{1-\delta_{mn}}{T_2}\rho_{mn}(\mathbf{k},t),        
    \end{split}
    \label{eq:EOM}
\end{equation}
Here $\epsilon_m(\mathbf{k})$ denotes the energy of the bands and $\mathbf{d}_{m,n}(\mathbf{k})$ the transition dipole matrix element between bands. We define the dipole to include the charge of the electron, so that in atomic units $\mathbf{d}_{mn}(\mathbf{k}) = -\langle u_m(\mathbf{k})|\mathbf{r}|u_n(\mathbf{k})\rangle$. This convention fixes the relative sign of the second term with respect to the interaction term of Eq. \ref{eq:Hamil} and is used consistently in the expressions for the currents below. The equation of motion contains four distinct terms. The first term, depending on the energy difference between the bands, describes the different phase accumulation between the carriers in the different bands. The second term, depending on the dipole coupling, describes the induced transition of carriers between bands. The third term describes the acceleration of the carrier within a band. The fourth term is a phenomenological addition which describes the loss of coherence between the different bands over time. The dephasing time $T_2$ implicitly describes all interactions that give rise to the loss of coherence, which are neglected in the Hamiltonian in Eq. \ref{eq:Hamil}. As HHG is a coherent emission process, the loss of coherence translates to a reduction of the generated harmonic intensity. In the simplified picture, we can think of a scattering event happening during the acceleration, which causes the electron-hole pair to lose their spatial coherence, inhibiting recombination, and thus their motion not contributing to harmonic emission, as shown in Fig. \ref{fig:TSM}c. 

Within reciprocal-space models that assume a uniform, $\mathbf{k}$-independent dephasing time, agreement with measured spectra is typically obtained for $T_2$ of the order of a few femtoseconds \cite{yue2022,Vampa2014a}. This should not be read as a measurement of the microscopic coherence time. Real-space calculations show that such a short uniform T$_2$ largely reproduces the effect of spatial and propagation-induced trajectory filtering, so that it acts as an effective parameter which absorbs mechanisms that may not be dephasing of electron-hole coherences, and which may in turn distort the microscopic and macroscopic response that is extracted \cite{Brown2024}. 

The dephasing times inferred from HHG simulations are shorter than those measured by femtosecond photon echo in comparable semiconductors \cite{Becker1988}. The comparison should be made with care, however. The echo measurements were performed at carrier densities well below those injected by a driving field strong enough for efficient above-bandgap HHG, and the same measurements already resolve a dependence of the dephasing rate on carrier density, which was assigned to carrier-carrier scattering \cite{Becker1988}. A single, intensity-independent $T_2$ is therefore an approximation of limited validity, and part of the apparent discrepancy may simply reflect the different excitation conditions. 
Various contributions to the dephasing have been studied, including temperature and phonon-scattering \cite{Du2022,Freeman2022,mokhtari2026,Cardenas2025}, and macroscopic propagation of the fields \cite{Floss2018,Kilen2020}. However, at the time of writing, the dominant contribution to the dephasing and the exact underlying mechanism is still uncertain.\\
From the equation of motion shown in Eq. \ref{eq:EOM} the total generated current during the HHG process can be evaluated, which can be separately expressed in the intraband and interband currents \cite{Vampa2017},
\begin{equation}
\begin{split}
    \mathbf{j}_{\text{tot}}(t) & = \mathbf{j}_{\text{intra}}(t)+\mathbf{j}_{\text{inter}}(t)\\
    \mathbf{j}_{\text{intra}}(t) & = -\int_{\text{BZ}}\sum_{m} \nabla_\mathbf{k}\epsilon_{m}(\mathbf{k})\rho_{mm}(\mathbf{k},t) \text{ d}^3\mathbf{k},
    \\
    \mathbf{j}_{\text{inter}}(t) & = \frac{\partial}{\partial t}\int_{\text{BZ}} \sum_{m\neq n}\mathbf{d}_{mn}(\mathbf{k})\rho_{nm}(\mathbf{k},t) \text{ d}^3\mathbf{k}.
    \\ 
\end{split}
\label{eq:currents}
\end{equation}
Both expressions are written in atomic units for an electron of charge $-1$. This charge is absorbed into the definition of $\mathbf{d}_{mn}$ given above, and appears explicitly as the minus sign of the intraband term; with this convention the prefactors $s_m$ introduced in Eq. \ref{eq:currents_analytic} below are consistent with Eq. \ref{eq:currents}. Note also that $\mathbf{j}_{\text{inter}}$ is already the time derivative of the interband polarization, $\mathbf{j}_{\text{inter}} = \partial \mathbf{P}_{\text{inter}}/\partial t$, so that both contributions to $\mathbf{j}_{\text{tot}}$ are currents and enter the emitted spectrum in the same way.\\

The intraband current arises from the motion of carriers within a band, and is therefore governed by the band dispersion and by the population distribution. The interband current follows from the coherences between bands. These coherences are not restricted to resonantly driven, real transitions: off-resonant driving generates interband coherence that leaves no residual population once the pulse has passed, which is what is termed a virtual transition in perturbative nonlinear optics. Expanding this contribution to lowest order in the field recovers the Kerr and higher-order non-resonant susceptibilities. The below-bandgap interband response is therefore contained in Eq. \ref{eq:currents} without introducing any additional or virtual bands, and both components can contribute below the band gap. Which of the two dominates there depends on the material and on the driving conditions rather than following a fixed ordering, whereas above the band gap the interband component generally carries most of the emission.\\

Throughout this chapter, interband and intraband denote the two components into which the calculated current is decomposed after projection onto the field-free Bloch states. They are not themselves microscopic mechanisms, and neither of them is uniquely associated with one physical process. Electron-hole recollision is one pathway that produces an interband polarization, but an interband polarization is also generated without any recollision, for instance by the non-resonant, Kerr-type response of the driven bands or by carriers that never return to their point of creation. Conversely, a nonlinear intraband current requires neither non-parabolic bands nor the absence of interband coupling. The two components are moreover dynamically coupled: the interband coherence feeds the populations that carry the intraband current, and the intraband motion sets the reciprocal-space position at which the interband coupling is evaluated. Since the decomposition depends on the choice of basis, and since only the total current is an observable, the identification of a single dominant microscopic mechanism from an experiment is correspondingly inhibited, and will rather rely on a model-based interpretation than a unique experimental observation.\\

Propagation of the driving and harmonic fields through the sample must be distinguished from microscopic dephasing. Phase mismatch, absorption and the spatial variation of the driving intensity select a subset of the emitted trajectories and can suppress the far-field signal in a way that resembles a short microscopic $T_2$, while the local electron-hole coherence remains intact. The two therefore cannot be separated on the basis of the harmonic yield alone.\\

The harmonic spectrum follows from the radiated field of this time-dependent current density. In the far field the emitted amplitude is proportional to the second derivative of the polarization, equivalently to the first derivative of the current, so that
\begin{equation}
    I(\omega) \propto \omega^2\Big|\mathbf{j}(\omega)\Big|^2 = \Bigg|\mathcal{F}\Big[\frac{\text{d}\mathbf{j}(t)}{\text{d}t}\Big](\omega)\Bigg|^2,
    \label{eq:spectrum}
\end{equation}
where $\mathcal{F}$ denotes the Fourier transform. The two forms are equivalent provided the current and its derivative vanish at both ends of the simulated time window and the calculation is converged with respect to the reciprocal-space sampling and the time step. In practice, we compute $\mathbf{j}(t)$, transform it, and apply the $\omega^2$ factor in the frequency domain, rather than differentiating in the time domain, as numerical differentiation amplifies high-frequency noise. In either case the current should be multiplied by a window function before transforming, and the finite dephasing time included in Eq. \ref{eq:EOM} already suppresses the tail of the signal, which reduces the sensitivity to the choice of window.\\

In the next part, we will illustrate how the semi-classical interpretation can be reconciled with the equation of motion shown above, see Eq. \ref{eq:EOM}. We consider a simplified system in which the material is described by two bands, a conduction band $\epsilon_c(\mathbf{k})$ and a valence band $\epsilon_v(\mathbf{k})$. In addition, we make use of the Keldysh approximation\cite{Keldysh1965}, in which the excitation fraction is considered to be small, so that the valence band remains close to fully occupied and the conduction population $\rho_{cc}(\mathbf{k},t)$ is retained only to lowest non-vanishing order in the field. This population is not set to zero, since it is precisely what carries the intraband current in the expression below. Finally, by moving from the length to the velocity gauge where the vector potential $\frac{\text{d}\mathbf{A}(t)}{\text{d}t} = -\mathbf{F}(t)$ and electric scalar potential $\Phi = 0$, the currents can be expressed analytically \cite{Vampa2017},
\begin{equation}
            \begin{split}
    \mathbf{j}_{\text{intra}}(\omega) = \sum_{m=c,v}& s_{m} \int_{-\infty}^\infty \Bigg(\int_{\text{BZ}} \int_{-\infty}^{t}  \int_{-\infty}^{t'} \mathbf{v}_m(\mathbf{k}) \Big[\mathbf{F}(t')\cdot\mathbf{d}^*\Big(\mathbf{k}+\mathbf{A}(t')-\mathbf{A}(t)\Big)\Big] \\
    & \cdot \Big[\mathbf{F}(t'')\cdot\mathbf{d}\Big(\mathbf{k}+\mathbf{A}(t'')-\mathbf{A}(t)\Big)\Big]\\
    & \cdot e^{iS(\mathbf{k},t'',t';t)-(t'-t'')/T_2}
    \text{ d}t'' \text{ d}t'\text{ d}^3\mathbf{k}  +\text{ c.c.}\Bigg)  e^{-i\omega t}\text{ d}t
    \\
    \mathbf{j}_{\text{inter}}(\omega) = \int_{-\infty}^\infty & \Bigg(\int_{\text{BZ}}  \int_{-\infty}^{t} \mathbf{d}^*(\mathbf{k})\Big[\mathbf{F}(t')\cdot\mathbf{d}\Big(\mathbf{k}+\mathbf{A}(t')-\mathbf{A}(t)\Big)\Big] \\
    & \cdot e^{iS(\mathbf{k},t',t;t)-(t-t')/T_2}
    \text{ d}t' \text{ d}^3\mathbf{k}  +\text{ c.c.}\Bigg)  e^{-i\omega t}\text{ d}t
    \\ 
    \end{split}
    \label{eq:currents_analytic}
\end{equation}
Note that the outer integral over $t$ carries the factor $e^{-i\omega t}$, so that the left-hand sides are frequency-domain quantities. Here $s_v=1$ and $s_c = -1$, which follows from Eq. \ref{eq:currents} once the electron charge is absorbed as described above, since the conduction electron and the valence hole contribute to the current with opposite sign. The band velocity is $\mathbf{v}_m(\mathbf{k}) = \nabla_\mathbf{k}\epsilon_m(\mathbf{k})$, and the semi-classical action is defined with an explicit reference time $t_{\text{ref}}$, at which the crystal momentum equals $\mathbf{k}$:
\begin{equation}
S(\mathbf{k},t_a,t_b;t_{\text{ref}})=\int_{t_a}^{t_b}\epsilon_{c,v}\Big(\mathbf{k}+\mathbf{A}(\tau)-\mathbf{A}(t_{\text{ref}})\Big) \text{d}\tau \quad \text{where} \quad \epsilon_{c,v}(\mathbf{k}) = \epsilon_c(\mathbf{k})-\epsilon_v(\mathbf{k}).
\label{eq:action}
\end{equation}
Both terms of Eq. \ref{eq:currents_analytic} use $t_{\text{ref}}=t$, consistently with the arguments of the transition dipoles.\\
We will consider a harmonic that has a dominant interband contribution. The generation process takes place approximately within an optical half-cycle of the driver, which fits a number of harmonic cycles, and thus by itself the $e^{-i\omega t}$ will average out to zero. The dominant contributions are those where the total phase stays stationary, which enables approximating the integral via the saddle point method. Writing the interband integrand as $e^{i\phi}$, the factor $e^{iS-(t-t')/T_2}e^{-i\omega t}$ gives
\begin{equation}
    \phi = S(\mathbf{k},t',t;t)-\omega t +\frac{i(t-t')}{T_2},
    \label{eq:saddlephase}
\end{equation}
where the sign of the dephasing term is fixed by requiring that the coherence decays rather than grows. Stationarity of $\phi$ with respect to $t'$, $\mathbf{k}$ and $t$ provides three distinct equations \cite{Vampa2017},
\begin{equation}
\begin{split}
    \frac{\text{d}\phi}{\text{d}t'} = 0 \quad &\rightarrow \quad \epsilon_{c,v}\Big(\mathbf{k}+\mathbf{A}(t')-\mathbf{A}(t)\Big)+\frac{i}{T_2} = 0,\\
    \nabla_\mathbf{k}\phi = 0 \quad &\rightarrow \quad \int_{t'}^t \mathbf{v}_{c,v}\Big(\mathbf{k}+\mathbf{A}(t'')-\mathbf{A}(t)\Big)\text{ d}t'' =0, \\
    \frac{\text{d}\phi}{\text{d}t} = 0 \quad &\rightarrow \quad \epsilon_{c,v}(\mathbf{k})-\omega+\frac{i}{T_2} = 0,
\end{split}
\label{eq:saddle}
\end{equation}
where the second condition has been used to drop the boundary term in $\text{d}\phi/\text{d}t$. With $\mathbf{v}_{c,v}(\mathbf{k}) = \mathbf{v}_c(\mathbf{k})-\mathbf{v}_v(\mathbf{k})$. The first equation has no real solution, as it implicitly describes the tunneling process. Neglecting dephasing and tunneling, the first condition defines a real birth momentum $\mathbf{k}_0$ through $\epsilon_{c,v}(\mathbf{k}_0)\rightarrow0$, and we find that $\mathbf{k} = \mathbf{k}_0+\mathbf{A}(t)-\mathbf{A}(t')$. The electron-hole pair created at the initialization time $t_i = t'$ at $\mathbf{k}_0$ will be driven through reciprocal space following the driver's net vector potential. The second equation denotes that the displacement between initialization and recombination $t_r = t$ should go to zero, or that the electron and hole have to spatially recombine. In the third equation, again neglecting dephasing, we find that the emitted energy is equal to the bandgap energy at the time of recombination, $\omega = \epsilon_{c,v}\Big(\mathbf{k}_0+\mathbf{A}(t_r)-\mathbf{A}(t_i)\Big)$. This allows us to interpret the harmonic emission as a sum of trajectories as conceptually described by the three-step model. The saddle-point approximation not only unites the three-step model with the SBEs but also enables semi-classical calculations of solid HHG where the electron-hole motion is evaluated classically.\\
The trajectory picture of HHG is not unique to solid HHG and was first developed in the context of gas HHG, where the approximations are more appropriate due to the lack of intraband current. In gases for each harmonic, two distinct trajectories are found: the short and long trajectories. The varying duration of the different trajectories gives rise to a different phase and divergence of the harmonics, which enables experimental selection of a specific trajectory. Contrary to gases, where the excitation is limited to a small region in momentum space, in solids, the excitation can occur all along the bands, which results in a more continuous distribution of potential trajectories \cite{Huttner2016}. 

\subsection{Time-Resolved High-Harmonic Generation}
Conventional solid HHG probes the equilibrium electronic structure through the nonlinear motion of carriers in the crystal. In transient HHG, an additional excitation pulse modifies the material prior to or during harmonic generation, causing the emitted harmonics to become sensitive to photoinduced changes in the electronic or structural state. Depending on the system and excitation conditions, these changes can manifest as variations in harmonic intensity, phase, polarization, spectral position, or through the appearance of wave-mixing signals. 

As the induced carrier motion in the material is greatly impacted by the material's potential landscape, modifications to the material are reflected in the harmonic emissions' amplitude and phase. Predicting these changes can be very challenging as precise modeling of the material is required. In this section, we will provide a few specific examples to illustrate how the harmonic emission can be modified.\\

Photoexcitation modifies the electronic structure of essentially any semiconductor, through bandgap renormalization, screening of the Coulomb interaction and band filling \cite{dekeijzer2024,VanDerGeest2023}. In conventional semiconductors these changes are nevertheless small on the scale of the gap itself, so that the band structure is perturbed rather than reorganized. Strongly correlated materials are the important exception, since there the gap is set by the interaction energy and can collapse entirely; these are discussed in a later section. These small energy shifts are reflected in the harmonic emission, specifically in the dipole phase, which is the spectral phase of the various harmonics relative to the driver \cite{Carlstrm2016,Lewenstein1995}. In the semi-classical framework, the dipole phase for a specific harmonic $q$ generated by a driver with frequency $\omega_0$ is given by \cite{Carlstrm2016}:
\begin{equation}
    \phi_d = q\Big(\omega_0 t_f - \frac{\pi}{2}\Big) -S(\mathbf{k}_0,t_i,t_f;t_f)
    \label{eq:dipolephase}
\end{equation}
The dominant contribution here is the semi-classical action, which we can interpret as the phase accumulated during the generation process. We can simplify our problem by considering a bandgap shift which is constant in reciprocal space $\epsilon_{c,v}(\mathbf{k})\xrightarrow{}\epsilon_{c,v}(\mathbf{k})+\Delta\epsilon$, providing us with a simple expression for the dipole phase shift \cite{dekeijzer2024},
\begin{equation}
    \Delta\phi_d \approx -\Delta\epsilon\, \Delta t \quad \text{with} \quad \Delta t = t_f-t_i,
    \label{eq:dipolephaseapprox}
\end{equation}
Here $\Delta t$ is the characteristic excursion time for a particular harmonic. The minus sign follows from the action entering Eq. \ref{eq:dipolephase} with a negative sign, since a constant gap shift changes the action by $\Delta S = \Delta\epsilon\,\Delta t$. A transient reduction of the band gap, as produced for instance by bandgap renormalization, therefore advances the dipole phase, while an increase retards it. Acquisition of the harmonic phase thus provides insight into energy shifts in the material. In the case where the bandgap modulation changes over time, such as in the case of coherent phonons, the time-dependent phase shifts can result in spectral shifts of the harmonic emission \cite{Zhang2024}.\\

The pre-excitation of the material can not only alter the band structure of the material but also affect the multi-particle interactions implicitly described by the dephasing time. This effect can generally be referred to as excitation-induced dephasing. As described above, increased dephasing manifests in the suppression of harmonics. In the semi-classical framework, the dephasing dependence of the harmonic intensity is found to scale exponentially \cite{dekeijzer2024},
\begin{equation}
    I \appropto e^{-\frac{\Delta t}{T_2}}\Big|T_2\Big(1-e^{-\Delta t/T_2}\Big)\Big|^2.
    \label{eq:dephasingscaling}
\end{equation}
This relation is not general. It follows from the same two-band, saddle-point treatment introduced above, in which a single dominant trajectory pair is retained for each harmonic order and the dephasing time is taken to be independent of $\mathbf{k}$. The first factor describes the damping of the amplitude of that trajectory, and the second the coherent sum over the emission times within the generation window. Within these assumptions the relation is in line with SBE simulation results \cite{dekeijzer2024} and experimental findings \cite{Heide2022}. Within the same two-band picture the excursion time grows with harmonic order, so that higher orders are more strongly affected by a change of the dephasing time, which is consistent with experiment. This ordering is not generic: where several bands or several trajectory families contribute, the excursion time associated with a given harmonic order need not increase monotonically with that order \cite{Huttner2016}, and the ordering of the suppression can then differ. As the energy of the excitation disperses throughout the material, the dephasing rate will be altered over time, making the harmonic intensity a probe for these characteristic lifetimes.\\

Several works have explored two-color HHG, where the driver and control beams are overlapped in time. Although their coupling still strongly depends on the material, most of these works focus on optimizing the temporal waveform of the driving field for HHG instead of probing the material. By combining a driver with its second or third harmonic, the temporal waveform can be controlled. Shaping the field so as to have maximum field strength at the initialization and recombination times increases the harmonic yield \cite{roscam2024,He2010,Burger2017,Kroh2018}. Alternatively, the use of incommensurate fields (where the control field is not a frequency multiple of the driver) causes harmonic suppression as the waveform is shaped to favor wavemixing over harmonic generation \cite{Wang2023,vanessen2026}.

\newpage
\section{Solid High-Harmonic Generation in an Experimental Setup}
In its simplest form, a solid-state HHG experiment consists of focusing a femtosecond pulse onto a solid sample and dispersing the emitted harmonics onto a detector. The practical requirements differ substantially depending on the highest harmonic order or highest photon energy of interest, and much of the relevant know-how is distributed across the method sections of individual papers. In this section we therefore describe the driving source and the constraints imposed by the sample, the generation geometry, the transition from ambient to vacuum operation, the collection, filtering and detection of the harmonics, and the main differences with respect to gas-phase HHG, including what is required to convert an existing gas-phase beamline. We close with the time-resolved implementation that underlies the measurements discussed in the remainder of this chapter.

\subsection{Driving Source and Sample Constraints}
The choice of driving wavelength is often a compromise. The ponderomotive energy and the reciprocal-space excursion both grow with wavelength at fixed field strength, which favors long wavelengths for reaching high harmonic orders, and the field strength required to reach a given cutoff drops accordingly. Most solid-state HHG experiments are therefore driven in the near- to mid-infrared, typically between roughly one and five micrometres, using optical parametric amplifiers or optical parametric chirped-pulse amplifiers pumped by Ti:sapphire or ytterbium-based systems. Pulse durations range from a few tens of femtoseconds to a few optical cycles, and pulse energies in the microjoule range are generally sufficient, since the required intensity is reached with a moderately tight focus. At the same time, HHG has been demonstrated with standard 800~nm Ti:sapphire pulses and even at shorter wavelengths, at the cost of a reduced cutoff \cite{luu2015,Luu2018e}.

The limiting constraint is not the available pulse energy but the damage threshold of the sample. Solid-state HHG operates at intensities approaching, but staying below, the threshold for optical breakdown, so that the accessible cutoff is bounded by the material rather than by the laser. This has three practical consequences. First, the repetition rate and the average power on the sample must be chosen such that heat accumulation does not lower the effective threshold. Second, damage is cumulative, so that samples are commonly translated or rastered during acquisition, and the harmonic yield should be monitored for a slow decay that signals incipient damage. Third, since the emission scales steeply with intensity, focus quality, pointing stability and pulse-energy stability translate directly into signal stability, and a reference measurement of the driving pulse energy on a shot-to-shot or shot-averaged basis is worthwhile.

\subsection{Generation Geometry}
Harmonics can be collected in transmission or in reflection \cite{Vampa2018}. In transmission the harmonics must propagate out of the sample, and the reabsorption length therefore sets the depth from which the detected signal originates. For photon energies above the band gap this length is short, of the order of tens of nanometres, so that the emission effectively originates from a thin layer at the exit surface and the sample thickness ceases to matter for the signal strength. Below the band gap the sample is transparent and the full interaction length contributes, which makes phase matching and driver reshaping relevant. In reflection the signal originates from the entrance surface, which makes the geometry the natural choice for opaque samples, for bulk crystals that are too thick to be used in transmission, and for samples on non-transmitting substrates. Reflection geometries also reduce the influence of propagation of the driver through the sample \cite{linden2025nonlinear}, at the cost of a smaller collection solid angle and of a stronger sensitivity to surface quality.

The driver is focused with a lens for near-infrared drivers or with an off-axis parabolic mirror for mid-infrared and few-cycle pulses, where chromatic dispersion and material absorption make refractive optics inconvenient. Numerical apertures are usually kept modest, so that the confocal parameter exceeds the generation depth and the intensity is well defined across the interaction region. Since the harmonic yield scales nonlinearly with intensity, the emission is confined to the central part of the focus, which is what makes point-scanning harmonic microscopy possible but also means that the effective source size is smaller than the driving focus.

\subsection{From Ambient to Vacuum Operation}
Harmonics in the infrared, visible and near-ultraviolet can be generated and detected under ambient conditions, which considerably simplifies the setup and allows conventional refractive optics and fiber-coupled spectrometers to be used. Around the vacuum-ultraviolet the situation changes, since molecular oxygen and water vapour absorb strongly, and from there on the beam path has to be evacuated or purged. For the extreme ultraviolet, windows are no longer available at all and the generation and detection chambers must form a single evacuated volume.

A useful practical point is that the vacuum requirements are far less demanding than for gas-phase HHG. There is no gas load from the target, so a pressure in the $10^{-5}$ to $10^{-6}$~mbar range is sufficient for transport and detection, and differential pumping between the generation and detection stages is generally unnecessary. What does require attention is contamination: hydrocarbon deposition on the sample surface under intense irradiation degrades both the damage threshold and the harmonic yield over time, so oil-free pumping and careful sample handling are advisable.

\subsection{Collection, Filtering, and Detection of the Harmonics}
In the extreme ultraviolet the harmonics cannot be collected or refocused with refractive optics, and normal-incidence reflectivity is low. Collection and recollimation are therefore performed either with grazing-incidence toroidal or ellipsoidal mirrors, which are broadband but introduce astigmatism away from the design conjugates, or with normal-incidence multilayer mirrors, which offer convenient geometry and good reflectivity at the price of a narrow bandwidth and thus the selection of a single harmonic order. In many experiments no collection optic is used at all, and the harmonics are allowed to propagate freely from the sample onto the spectrometer entrance, which preserves the divergence information that carries the harmonic wavefront.

Separating the harmonics from the driving field is more demanding than in a gas jet, since the sample sits in the beam and the harmonics are emitted collinearly with the transmitted or reflected driver, which is many orders of magnitude more intense. Thin metallic foils of a few hundred nanometres thickness are the standard solution in the extreme ultraviolet, with aluminium, zirconium, tin and indium each transmitting a characteristic band while blocking the infrared completely. In the vacuum- and near-ultraviolet, dielectric mirrors, colored-glass filters and the dispersion of the spectrometer itself take over this role. A geometric contribution to the separation can be obtained by introducing a small angle between the generation and control beams, which sends the various harmonics and wave-mixing orders into distinct directions \cite{roscam2024,vimal2023photon,Zhang2025,Luttmann2023,Bertrand2011a,Heyl2014}.

For detection, the extreme ultraviolet is most commonly dispersed by a flat-field grazing-incidence spectrometer with a variable-line-spacing grating, and recorded either with a microchannel plate followed by a phosphor screen and a camera, or directly with a back-illuminated CCD or CMOS sensor. The microchannel plate provides gain and is well suited to photon-starved measurements, while direct detection offers better linearity and a straightforward photon calibration. Both preserve one spatial dimension, so that the divergence of each harmonic order is recorded together with its spectrum, which is valuable for separating microscopic from propagation effects. For the low orders in the visible and ultraviolet, fiber-coupled grating spectrometers with CCD detection are adequate, and for single-order measurements a photomultiplier or a photodiode combined with a mechanical chopper and lock-in detection gives excellent sensitivity to the small relative modulations that are the observable in time-resolved experiments.

\subsection{Differences from Gas-Phase HHG and Adaptation of an Existing Beamline}
Several of the points above follow from one structural difference: the generating medium is a solid of sub-micrometre effective thickness rather than an extended gas. This removes the gas load and the associated differential pumping, but it also removes phase matching as a tuning parameter, since the coherence length is generally longer than the effective generation depth for above-bandgap harmonics. Absorption within the medium replaces phase matching as the quantity that determines how much of the sample contributes. The sample itself is in the beam and is consumed by the interaction, which has no counterpart in a gas target and makes translation stages, damage monitoring and sample exchange part of the standard experiment. Conversely, the solid can be structured, patterned and locally modified, which is the basis of the applications discussed in Section 6 and has no gas-phase analogue.

For a group with an existing gas-phase HHG beamline, the conversion is in most cases straightforward. The gas cell or jet is replaced by the sample mounted on a vacuum-compatible translation stage at the focus, with sufficient travel for rastering and, where required, with cooling or heating. The extreme-ultraviolet spectrometer and detection chain can be retained unchanged. The differential pumping stage becomes unnecessary, although leaving it in place is harmless. Metallic filters are required as before, but their role is now to block a driver that is transmitted or reflected by the sample rather than one that has diverged away from a gas jet, so their damage threshold should be checked. The driving source is the main investment, since the mid-infrared wavelengths favorable for solid-state HHG are not usually available on a gas-phase beamline. Finally, a pump arm with a delay stage has to be added for the time-resolved measurements described next.

\subsection{Time-Resolved Implementation}
\begin{figure}[H]
    \centering
    \includegraphics[width=\linewidth]{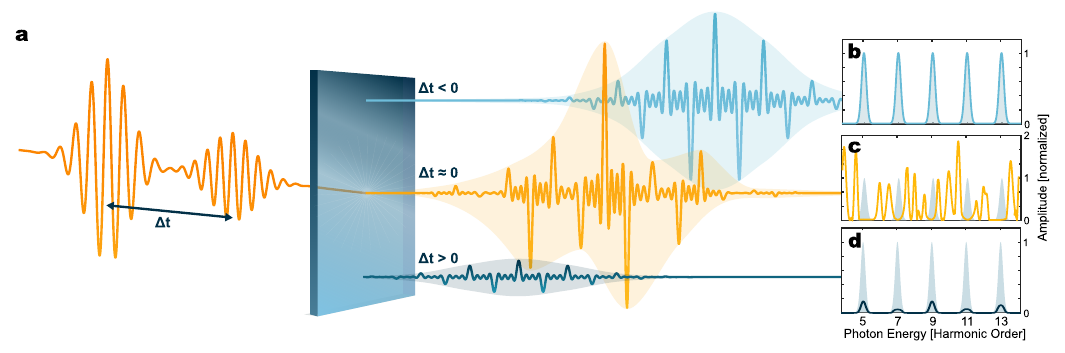}
    \caption{Schematic illustration of time-resolved high-harmonic generation from a solid sample. (a) Generation and control beams focused onto the sample, with the harmonics dispersed onto a detector. (b-d) Schematic HHG spectra for different delays. 
    (b) An unperturbed harmonic spectrum. 
    (c) A spectrum during temporal overlap showing clear wavemixing. 
    (d) A spectrum showing persistent suppression. 
    The shaded areas indicate the unperturbed harmonic spectra. Adapted from ref \citenum{vanessen2026}.}
    \label{fig:schematic}
\end{figure}

Time-resolved measurements are performed in a pump-probe geometry in which an additional femtosecond control pulse is introduced, as illustrated in Fig. \ref{fig:schematic}. If the control pulse precedes the generation pulse, it can excite carriers or phonons, and the generated harmonics serve as a probe of the photoinduced dynamics. During the temporal overlap of both pulses, wave mixing is enabled, which can be thought of as non-degenerate high-harmonic generation in which both pulses contribute $(n,m)$ photons to the wave-mixing signal. The control pulse can either be collinear with the generation pulse or be introduced at an angle, the latter separating the various harmonics and wave-mixing signals in emission angle \cite{roscam2024,vimal2023photon,Zhang2025,Luttmann2023,Bertrand2011a,Heyl2014}. Since the observable is usually a relative change of the harmonic yield of a few percent, chopping the control beam and referencing consecutive shots is the standard way to reach the required sensitivity, and the pump fluence should be kept in a range where the induced modulation is reversible and the sample recovers between shots. Besides measuring the intensity, wavelength, and divergence of the harmonics, a number of works have additionally measured the harmonic phase using various interferometric techniques \cite{koll2025extreme,Kuzkova2026,Lu2019,Kim2019,guery2026}.

\section{High-Harmonic Generation Spectroscopy of Semiconductors}
Solid HHG was first experimentally observed in bulk crystalline ZnO, but has since then been observed in various solids, including conventional semiconductors \cite{ghimire2011,Schubert2014,Vampa2015a,You2017,Vampa2015b}, thin films \cite{VanDerGeest2023}, two-dimensional materials \cite{Liu2017,Yoshikawa2017}, strongly correlated electron materials \cite{Bionta2021,Nie2023,uchida2022ruthenate}, epsilon-near-zero materials,\cite{yang2019high} and metals \cite{Korobenko2021,gholammirzaei2025highharmonicgenerationnoble}. In this section, we focus on the experimental findings from semiconductors. As discussed in the previous section, the harmonic emission is sensitive to a variety of microscopic quantities, including the band structure, transition dipole moments, Berry phases, carrier coherence, and scattering. A major effort in semiconductor HHG has therefore been to establish which material properties are encoded in the harmonic emission and how they can be extracted experimentally. The examples discussed below illustrate how HHG can be used to probe both equilibrium electronic structure and transient dynamics.

The first four subsections focus on work and general principles that highlight the sensitivity of HHG to a broad range of material properties. These include the laser-driven sub-cycle electron dynamics during HHG and the attochirp (Sec. \ref{sec:attochirp}), the electronic structure of the generation medium (Sec. \ref{sec:bandstructure}), the crystal structure and symmetry (Sec. \ref{sec:symmetry}), and the effects of reduced dimensionality (Sec. \ref{sec:confinement}). Together, these studies demonstrate the potential of HHG to probe material properties that are difficult to access by conventional spectroscopy. At the same time, many spectroscopic signatures cannot be linked unambiguously to a single microscopic origin. While this complicates interpretation and often requires multidimensional measurement approaches, it also implies that harmonic emission can be strongly modified by changes to the material or driving field. This sensitivity forms the basis for controlling and programming HHG, which is discussed in the final subsection (Sec. \ref{sec:intensitymodulation}). 

\subsection{Sub-Cycle Electron Dynamics and Attochirp}
\label{sec:attochirp}

Initial experimental efforts targeted the question of which of the two current components dominates the emission, and under which conditions. According to the simplified trajectory picture outlined above, intraband emission is expected to be nearly chirp-free, whereas interband emission can exhibit a negative chirp, as the different electron trajectories that give rise to different emitted photon energies spend a distinct time in the continuum. Consequently, one of the first spectroscopic observables investigated was the temporal structure of the harmonic emission itself. The temporal phase of solid HHG was characterized via streaking\cite{Garg2016} and two-colour approaches,\cite{Vampa2015b} with the goal of linking the observed attochirp to the underlying generation mechanism.

The subcycle temporal structure measurements of HHG in Quartz show no chirp,\cite{Garg2016} which was interpreted as the signature of intraband Bloch oscillation, whereas a significant chirp has been observed in HHG from ZnO crystal and was mainly assigned to interband polarization.\cite{Vampa2015b} Later experimental and theoretical studies suggested a more nuanced link between the temporal structure of HHG and the underlying mechanism. It was shown that the chirp in solid HHG is a joint characteristic of quantum interference involving multiple transitions in the band structure, and that even interband mechanisms can exhibit no chirp in a multiband system.\cite{Hohenleutner2015,Huttner2016} Particularly, chirp in an interband-dominated emission was shown to arise if only two bands are involved in the process, the bands follow a parabolic structure, and excitation is localized to one point in k-space.\cite{Huttner2016} Otherwise, also interband-dominated HHG can exhibit chirp-free emission.

These findings illustrate that the temporal structure of the harmonic emission is not merely a signature of whether HHG is interband or intraband in nature, but also reflects the details of the participating electronic structure and quantum pathways. For non-perturbative harmonics, a partial consensus has emerged that the intraband component carries most of the emission below the minimum band gap and the interband component most of the emission above it.\cite{Ghimire2019} The statement is restricted to that regime. Below the gap the interband component does not vanish, since the non-resonant interband coherence discussed in Section 2 contributes there as well, and the balance between the two depends on the material and on the driving conditions. A complementary view of the same sub-cycle physics is obtained without invoking a recollision picture at all, by reconstructing the sub-cycle ionization and carrier-injection dynamics from the low-order harmonic response of a driven dielectric.\cite{Juergens2022,juergens2024linking} Beyond clarifying the microscopic origin of HHG, these studies demonstrated the possibility of controlling the temporal waveform of the emitted harmonics and paved the route towards high-frequency (up to Terahertz) electronics.\cite{borsch2023lightwave}

\subsection{Band Structure, Density of States, and Transition Dipole Moment}
\label{sec:bandstructure}

The temporal and spectroscopic phases of solid HHG reconstructed from such time-domain measurements discussed in the previous subsection also reveal microscopic properties of the solid. In the microscopic picture of HHG, quantities such as the band structure, Berry phase, density of states, and transition dipole moment directly enter the equation of motion. Consequently, these quantities can become encoded in the harmonic emission and, under suitable conditions, be reconstructed experimentally.

\begin{figure}[H]
\centering
\includegraphics[width=0.8\linewidth]{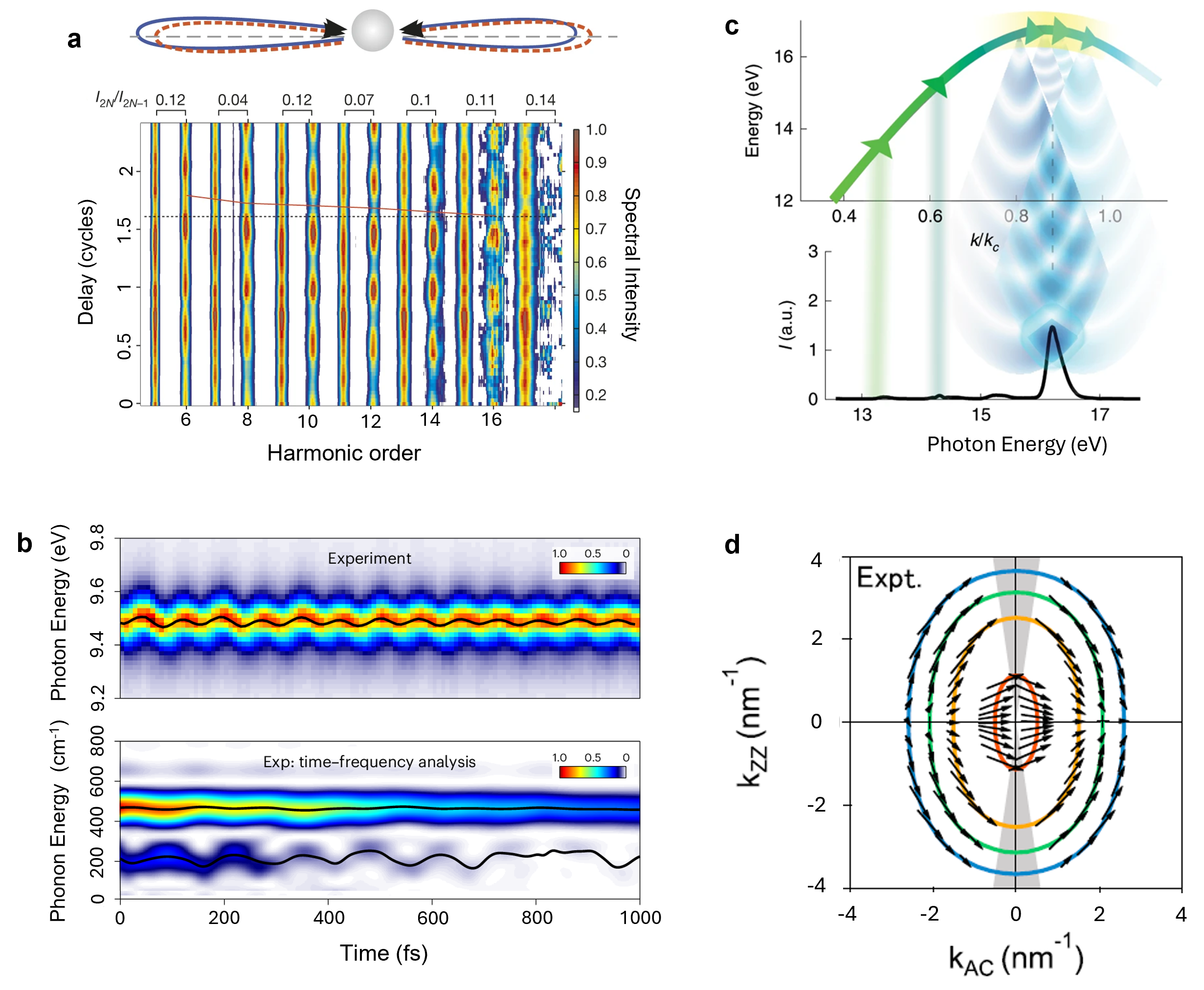}
\caption{Spectroscopic information extracted from solid HHG: (a) the quantum interference picture of solid HHG, and the delay phase of the even-order signals as a function of harmonic orders. (b) transient oscillation of HHG photon energy, driven by coherent phonon in the upper panel, and transient oscillation frequency analysis showing two phonon modes in the lower panel. (c) The enhancement of a certain harmonic order in MgO, related to the singularity of JDOS in momentum space. (d) Two-dimensional transition dipole moment reconstruction from experimental anisotropic solid HHG measurements.     
    (a) Adapted with permission from ref \citenum{Vampa2015b}. Copyright 2015 Springer Nature.
    (b) Adapted with permission from ref \citenum{Zhang2024}. Copyright 2024 Springer Nature.
    (c) Adapted with permission from ref \citenum{Uzan2020}. Copyright 2020 Springer Nature.
    (d) Adapted with permission from ref \citenum{uchida2021visualization}. Copyright 2021 American Physical Society.
}
\label{fig:BS}
\end{figure}

One illustration is shown in Fig.~\ref{fig:BS}a \cite{Vampa2015b}. HHG in both gases and solids results from the interference of quantum paths, and the joint interaction between the temporal symmetry of the driving field and the inversion symmetry of the lattice prohibits even-order harmonic generation. The introduction of a second field, for example, the second harmonic of the driver, transiently breaks this symmetry and enables even-order harmonics. The intensity modulation of these even harmonics depends on the harmonic order and reflects the coherent motion of the electron-hole pair. Through analysing this phase delay in ZnO, Vampa \emph{et al.} reconstructed the momentum-dependent band structure and identified split-off valence bands.\cite{Vampa2015a}

The same two-colour approach has subsequently been extended to probe geometric phases associated with both intraband and interband motion.\cite{bai2024probing,uzan2024observation} The sensitivity of HHG to the electronic structure is further illustrated by the coherent phonon measurements shown in Fig.~\ref{fig:BS}b.\cite{Zhang2024} Here, coherent lattice motion transiently modifies the band structure and thereby modulates the emitted harmonic photon energies. Through frequency analysis of these oscillations, Zhang \emph{et al.} extracted two phonon modes with different lifetimes and coupling strengths. This demonstrates how transient HHG can be used to probe electron-phonon coupling through its influence on the electronic structure.

Besides the band structure itself, HHG is also sensitive to the joint density of states (JDOS) and the transition dipole moment (TDM), both of which also play a central role in the semiconductor Bloch equations. 
When studying anisotropic solid HHG in MgO, Uzan \emph{et al.} observed a strong enhancement of specific harmonic orders at particular crystal orientations that could not be explained by the classical trajectory model.\cite{Uzan2020} The enhancement was attributed to spectral caustics arising from constructive quantum interference when the group velocity approaches zero near a Van Hove singularity, as illustrated in Fig.~\ref{fig:BS}c. Subsequent theoretical work further clarified the influence of the dynamical JDOS and its mapping onto the HHG spectrum.
Furthermore, using a one-dimensional two-band model of ZnO, Jiang \emph{et al.} showed that the amplitude and phase of the TDM predominantly affect odd and even harmonics, respectively.\cite{jiang2018role} Additionally, Uchida \emph{et al.} experimentally reconstructed the in-plane TDM texture in black phosphorus by measuring the parallel and perpendicular components of anisotropic solid HHG.\cite{uchida2021visualization} Fig. ~\ref{fig:BS}d shows the resulting TDM texture, where the coloured lines indicate the isoenergy contours associated with different harmonic orders and the black arrows represent the normalized TDM vectors.

These examples illustrate that HHG is sensitive to a broad range of microscopic quantities. Besides the bandgap itself, semiconductor HHG can provide information on band dispersion, geometric phases, electron-phonon coupling, density of states, and transition dipole moments, all of which are encoded in different observables of the harmonic emission.

\subsection{Lattice Structure and Symmetry}
\label{sec:symmetry}
In solids, the fixed crystalline lattice makes the angular dependence of HHG directly accessible without the need for molecular alignment. The harmonic emission can therefore carry information about crystal symmetry, real-space electronic structure, Berry phases, and momentum-dependent transition pathways. Anisotropic HHG has therefore emerged as a powerful spectroscopic tool for probing crystal symmetry, Berry phases, transition pathways, and real-space electronic structure.

One of the earliest examples was the observation of Berry-phase effects in angular HHG measurements from monolayer \ce{MoS2}.\cite{Liu2017} Later, a real-space semi-classical model based on recollisions with neighbouring atomic sites was proposed to describe the anisotropic solid HHG observed in MgO and successfully explained the emergence of additional anisotropic components with increasing laser intensity.\cite{You2017,juergens2024linking} Furthermore, Silva \emph{et al.} proposed describing HHG in a local Wannier basis rather than the conventional Bloch basis.\cite{Silva2019} Within such a real-space picture, anisotropic HHG becomes naturally linked to atomic potentials and chemical bonding, making it particularly attractive as a probe of local electronic structure.\cite{morimoto2021asymmetric}

\begin{figure}[H]
\centering
\includegraphics[width=0.8\linewidth]{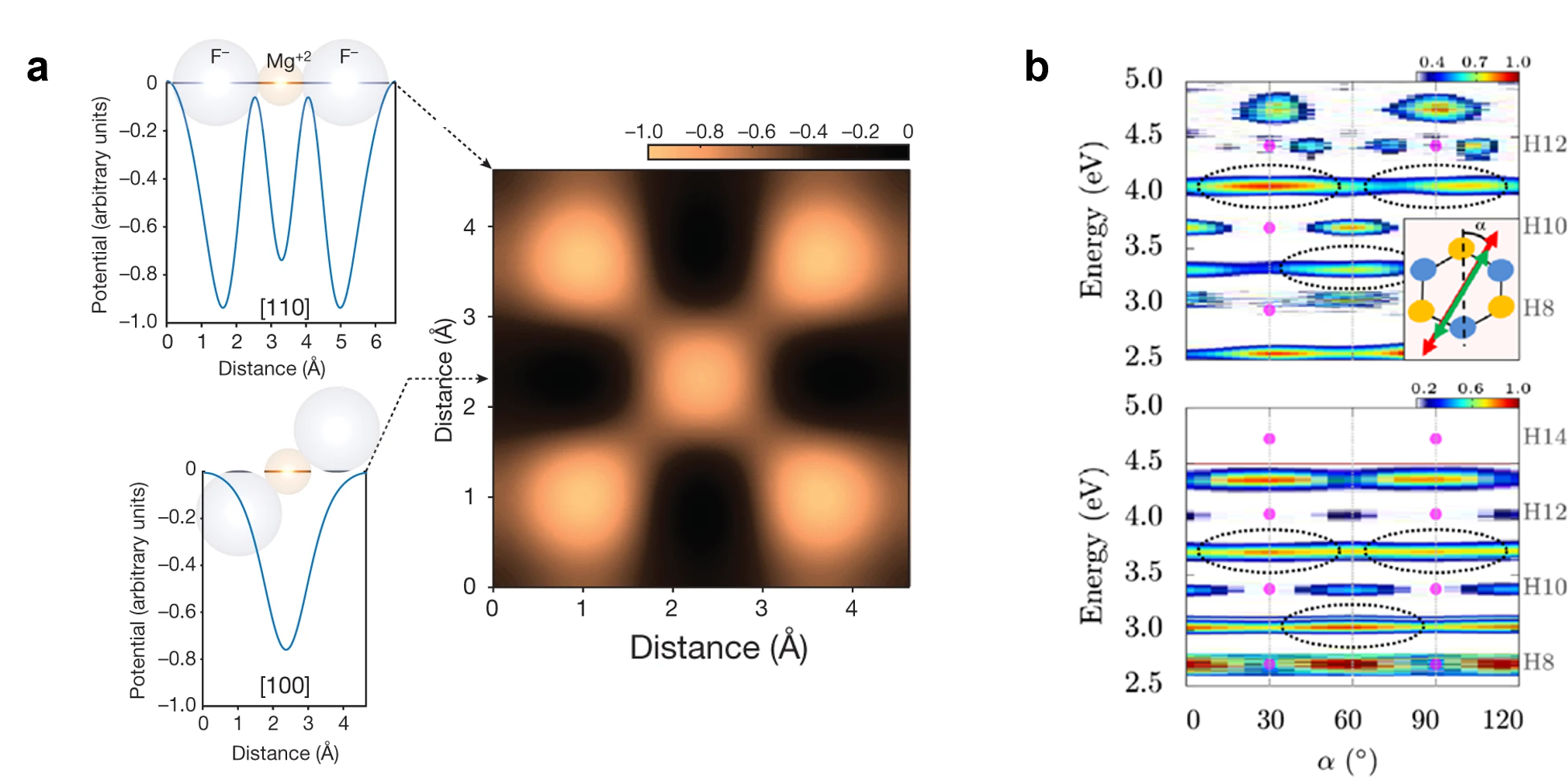}
\caption{(a) A picometer-scale potential mapping in \ce{MgF2} via anisotropic solid HHG measurements. (b) The anisotropic HHG spectrum in monolayer \ce{MoS2} with two different drive wavelengths.
    (a) Adapted with permission from ref \citenum{lakhotia2020laser}. Copyright 2020 Springer Nature.
    (b) Adapted from ref \citenum{yue2022signatures}. Copyright 2022 American Physical Society.
}
\label{fig:Anisotropy}
\end{figure}

An example of this capability is shown in Fig.~\ref{fig:Anisotropy}a, where Lakhotia \emph{et al.} demonstrated picometre-scale mapping of the valence potential and electron density in \ce{MgF2}.\cite{lakhotia2020laser} The angular dependence of HHG can also reveal information about multiband dynamics. Early SBE descriptions often relied on simplified two-band models, whereas the role of multiple conduction bands remained difficult to access experimentally. Yue \emph{et al.} observed that the angular dependence of HHG in monolayer \ce{MoS2} varies strongly with harmonic order, while remaining pinned around characteristic photon energies, as shown in Fig.~\ref{fig:Anisotropy}b.\cite{yue2022signatures} This behaviour was attributed to the presence of multiple recombination pathways involving different conduction bands, each possessing a distinct anisotropic transition strength.

The sensitivity of anisotropic HHG to the electronic structure has subsequently been exploited to probe transient and externally induced modifications of the band structure. Using a combination of anisotropic HHG and two-colour spectroscopy, laser-induced merging of adjacent conduction bands was observed through changes in the anisotropic phase response.\cite{uzan2024observation} More recently, the same sensitivity has been used in reverse to probe structural distortions under high pressure, demonstrating the applicability of anisotropic HHG as a probe of lattice-driven electronic changes.\cite{nebgen2026isostructural}

Overall, anisotropic HHG provides information that is complementary to the spectroscopic observables discussed above. While harmonic energies and phases primarily probe the electronic structure, the angular dependence of the emission additionally reveals information about crystal symmetry, real-space electronic structure, and multiband transition pathways.

\subsection{Quantum Confinement}
\label{sec:confinement}

Reducing the dimensionality of a material modifies both its electronic structure and optical response. As a result, HHG from low-dimensional systems can differ substantially from that of bulk semiconductors, and vice versa HHG provides a probe of extreme spatial scales. Quantum confinement changes the density of states, modifies transition energies, enhances Coulomb interactions, and can introduce entirely new symmetry properties. Consequently, HHG has become a useful tool for studying how strong-field dynamics evolve from bulk crystals to atomically thin materials and quantum-confined systems.

\begin{figure}[H]
\centering
\includegraphics[width=0.8\linewidth]{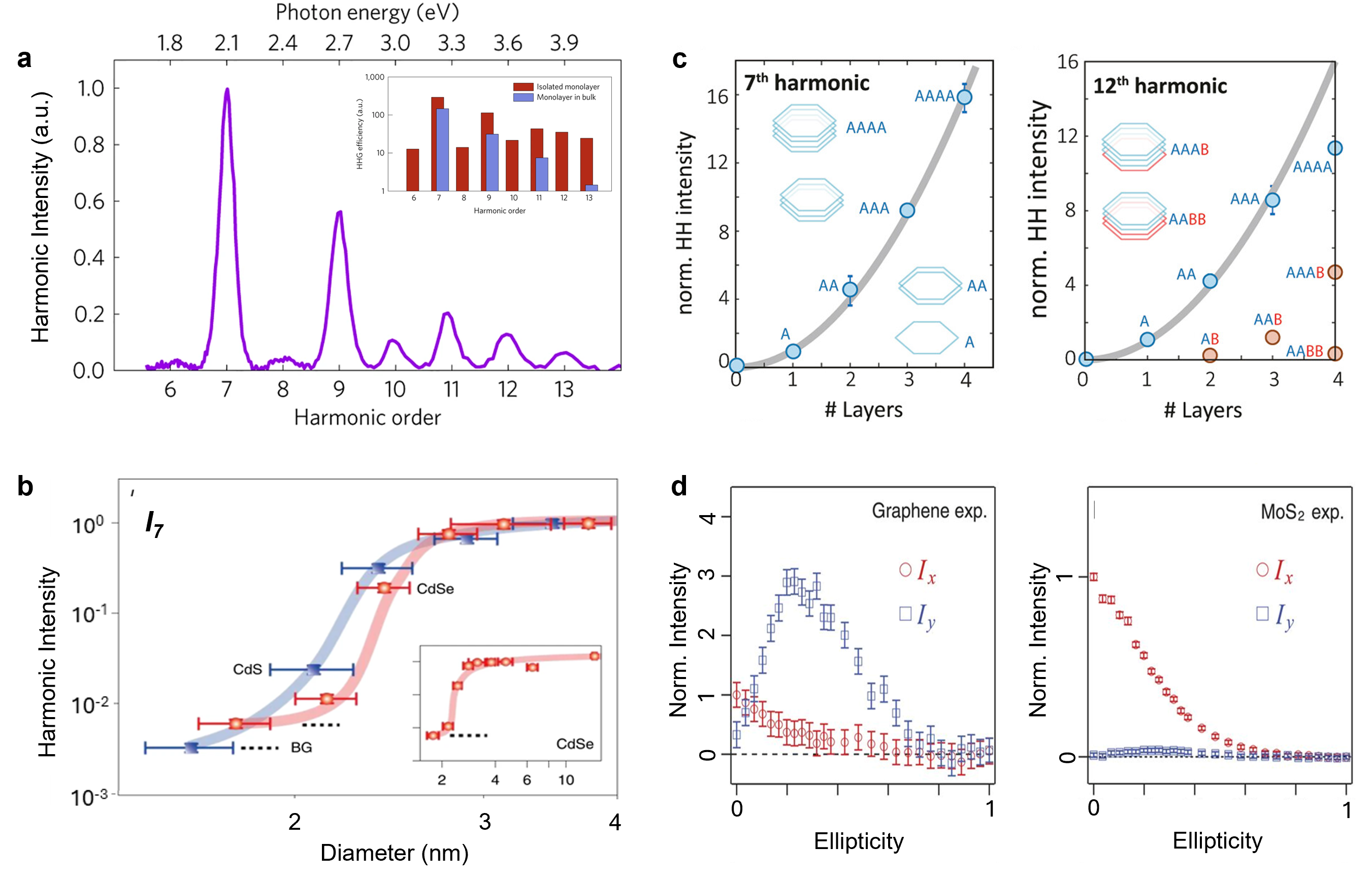}
\caption{Extraordinary HHG from low-dimensional semiconductors: (a) A typical HHG spectrum from monolayer \ce{MoS2}, and the inset shows the efficiency comparison between monolayer and bulk (b) The step-like dependence of HHG efficiency on the quantum dot size. 
(c) solid HHG intensity variation of artificially stacked 2D materials, as a function of layers and stacking symmetry.
(d) Anomalous elliptical HHG performance in graphene, compared with that in \ce{MoS2}, 
    (a) Adapted with permission from ref \citenum{Liu2017}. Copyright 2017 Springer Nature.
    (b) Adapted with permission from ref \citenum{nakagawa2022size}. Copyright 2022 Springer Nature.
    (c) Adapted from ref \citenum{heide2023high}. Copyright 2023 De Gruyter under CC BY 4.0, https://creativecommons.org/licenses/by/4.0/.
    (d) Adapted with permission from ref \citenum{Yoshikawa2017}. Copyright 2017 American Association for the Advancement of Science.
    }
\label{fig:confinement}
\end{figure}

One prominent example is provided by monolayer transition metal dichalcogenides. Quantum confinement in these atomically thin semiconductors is known to enhance perturbative nonlinear optical processes such as second-harmonic generation, and a similar enhancement has also been observed in HHG. As shown in Fig.~\ref{fig:confinement}a, monolayer \ce{MoS2} exhibits nearly an order-of-magnitude enhancement of the harmonic yield compared to the bulk material.\cite{Liu2017}

An even stronger confinement regime is reached in quantum dots. Fig. ~\ref{fig:confinement}b shows the harmonic efficiency measured from CdS and CdSe quantum dots as a function of particle size.\cite{nakagawa2022size} In both systems, the HHG efficiency exhibits a step-like dependence on the dot size. This behaviour was attributed to confinement-induced modifications of the sub-bandgap energy, which alter both intraband and interband transitions and thereby influence the HHG efficiency.

Reduced dimensionality also provides opportunities to engineer symmetry in ways that are not accessible in bulk materials. Artificially stacked two-dimensional materials allow the layer number and relative crystal orientation to be controlled independently. As shown in Fig.~\ref{fig:confinement}c, HHG from aligned stacks exhibits enhanced emission due to coherent in-phase interference, whereas twisted structures show a suppression of the harmonic yield caused by symmetry breaking.\cite{heide2023high} Such systems provide a controllable platform for studying the relationship between crystal symmetry and harmonic generation.

A particularly striking example is graphene. Most semiconductors exhibit their maximum HHG efficiency under linearly polarized excitation. In contrast, HHG in graphene reaches its maximum for a finite ellipticity of approximately 0.2, as shown in Fig.~\ref{fig:confinement}d.\cite{Yoshikawa2017} This unusual behaviour is absent in few-layer graphene and has been attributed to the Dirac-like band structure and massless carriers of the monolayer system.

These examples illustrate that quantum confinement does not only modify the efficiency of HHG, but can fundamentally alter the underlying strong-field dynamics. As a result, low-dimensional materials provide a valuable platform for investigating how electronic structure, symmetry, and dimensionality influence HHG.

\subsection{Intensity Modulation in Pump-Probe and Multicolor HHG}
\label{sec:intensitymodulation}
\begin{figure}
    \centering
    \includegraphics[width=\linewidth]{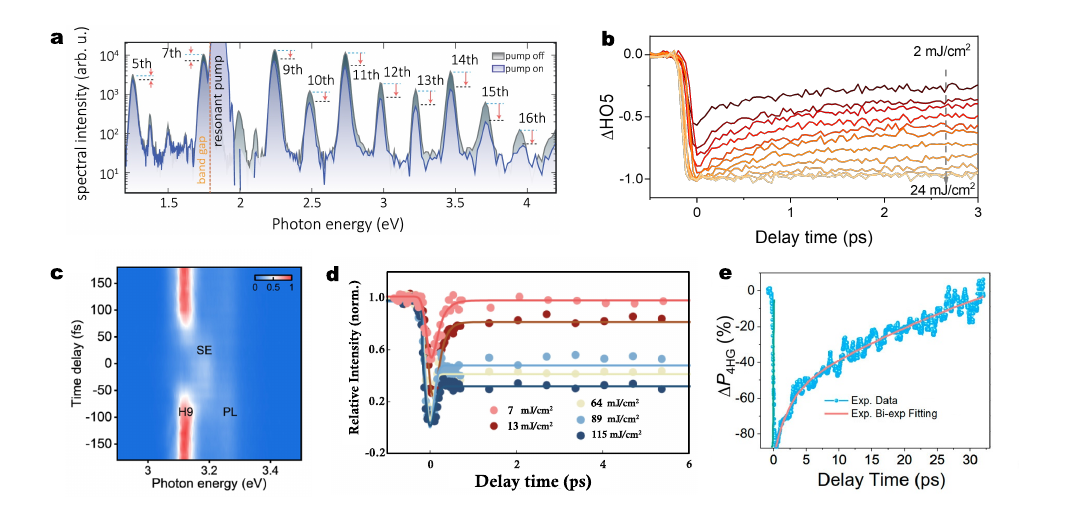}
    \caption{
    Time-resolved HHG results from various materials showing suppression. In (a) harmonic spectra from MoS$_2$ generated by 2000 nm and using a 660 nm control beam. The two different spectra are respectively with and without the control beam. The delay between the pulses is about 1 ps. In (b) the integrated fifth harmonic intensity generated by 2000 nm from NbO$_2$ and using a 400 nm control beam. From light to dark for increasing control fluence. In (c) the spectral delay map zoomed in on the ninth harmonic generated by 3500 nm from ZnO and modulated by 1300 nm. SE and PL are respectively indicating stimulated emission and photoluminescence signals. In (d) the integrated fifth harmonic generated by 2350 nm from ZnO and modulated by 800 nm. The legend indicates the different control fluences. In (e) the integrated fourth harmonic intensity generated by 1740 nm from a MoS$_2$ monolayer and modulated by 400 nm.
    (a) Adapted from ref \citenum{Heide2022}. Copyright 2022 Optica Publishing Group under CC BY 4.0, https://creativecommons.org/licenses/by/4.0/.
    (b) Adapted from ref \citenum{Nie2023}. Copyright 2023 American Physical Society under CC BY 4.0, https://creativecommons.org/licenses/by/4.0/.
    (c) Adapted with permission from ref \citenum{Wang2023}. Copyright 2023 American Physical Society.
    (d) Adapted from ref \citenum{Xu2022}. Copyright 2022 Optica Publishing Group under CC BY 4.0, https://creativecommons.org/licenses/by/4.0/.
    (e) Adapted from ref \citenum{Wang2022}. Copyright 2022 American Chemical Society under CC BY 4.0, https://creativecommons.org/licenses/by/4.0/.
    }
    \label{fig:control_of_HHG}
\end{figure}

\begin{figure}[H]
    \centering
    \includegraphics[width=\linewidth]{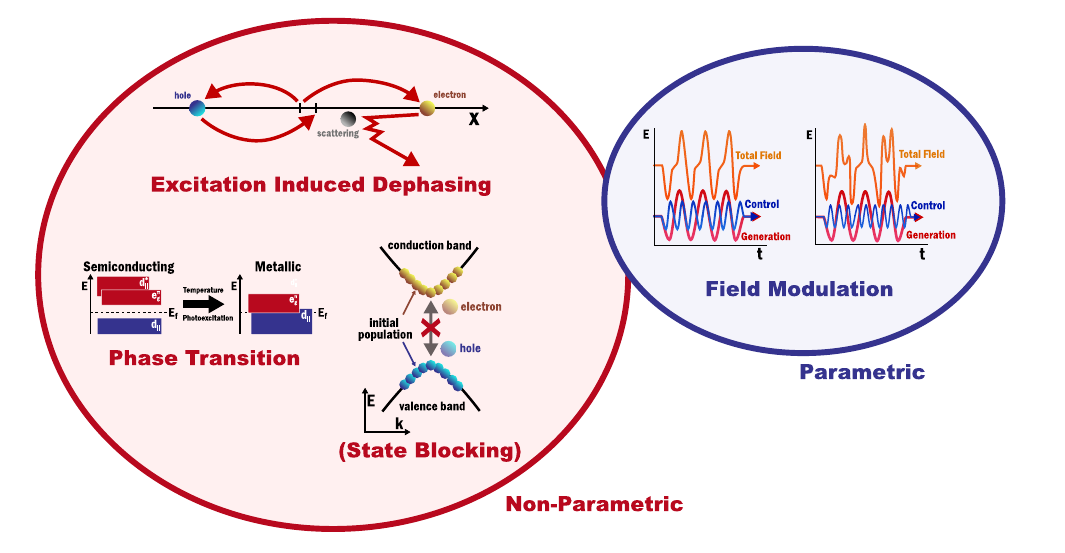}
    \caption{An overview of the microscopic mechanisms underlying harmonic suppression. State blocking is shown in brackets, as it is shown to be a minor effect in conventional semiconductors. Adapted from ref \citenum{essen2024HHGcontrol}. Copyright 2024 American Chemical Society under CC BY 4.0, https://creativecommons.org/licenses/by/4.0/.}
    \label{fig:control_mechanisms}
\end{figure}

Intensity modulation is one of the key observables in time-resolved solid HHG experiments. Although a few works have observed long-lived enhancement \cite{nie2024zno,Xu2023,Suthar2024}, the majority of works have reported a reduction of the harmonic yield upon the introduction of the control beam \cite{essen2024HHGcontrol}. A handful of exemplary results is shown in Fig. \ref{fig:control_of_HHG}a-e, showing harmonic suppression in a wide variety of platforms. 
The consistent suppression points towards general suppression mechanisms in TR-HHG and illustrates the potential of HHG as a universal, all-optically switchable process that can be actively exploited.\\
The suppression dynamics have been attributed to a variety of microscopic mechanisms, as schematically illustrated in Fig. \ref{fig:control_mechanisms}. The suppression can arise from the direct modulation of the HHG process by the introduction of an additional field, which shapes the temporal waveform. For this field modulation, temporal overlap of both beams is required, and the process is parametric, meaning that no energy is transferred to the material by the control beam. The parametric nature enables switching frequencies only limited by the duration of the pulses used. This means that even for moderate pulse durations of 100 fs the potential switching frequency is about 10 THz. In materials where other mechanisms are minimal, this leads to suppression restricted to temporal overlap as shown in Fig. \ref{fig:control_of_HHG}c.\\
Secondly, the suppression can arise from an alteration to the material induced by the pump, which is, by definition, non-parametric. In this case, harmonic suppression persists beyond temporal overlap, slowly recovering over time with lifetimes depending on the induced transient material alteration, as can be seen in Fig \ref{fig:control_of_HHG}b,d,e. Three distinct material alterations were considered to enable suppression. \\
Initial works suggested that the suppression might be caused by state blocking, where the pre-excitation of carriers by the pump blocks the generation of coherent electron-hole pairs, thus reducing the HHG yield \cite{Wang2017,Wang2022,Cheng2020}. Simulations, however, suggest that the effect of state blocking in most systems is minimal, as very high carrier excitation fractions of tens of percent are required to reach the amount of suppression observed experimentally \cite{Wang2017,Nagai2023}. At these high excitation fractions, most materials would reach dielectric breakdown and become damaged.
More recent works have instead attributed excitation-induced dephasing, where the pre-excitation by the control beam causes an increase in the dephasing rate, to be the main mechanism underpinning the long-lived harmonic suppression \cite{Heide2022,Xu2022,Nagai2023,VanDerGeest2023}. Simulations have illustrated that small changes to the dephasing time greatly affect the harmonic yield, which allows excitation-induced dephasing to explain the levels of suppression observed \cite{dekeijzer2024,Heide2022,Nagai2023}. In addition, the observation of increased suppression for higher-order harmonic matches with this underlying mechanism, as the higher-order harmonics have a longer associated excursion time. The recovery of the harmonic yield can happen over time scales of hundreds of femtoseconds to multiple picoseconds and longer, depending on the excited state lifetimes. 
Lastly, some of the strong HHG suppression observed in strongly correlated electron materials could directly be linked to the material phase transition, which we will discuss in the next section. 
Note that these mechanisms all consider the microscopic underlying principle of the HHG process and not the macroscopic propagation. As HHG is a coherent generation process, it is subject to phase-matching effects. Non-uniform modulation in a finite-size solid sample can modulate the effective phase-matching, which will be reflected in the harmonic intensity as well. The prominence of this effect depends on the effective depth over which harmonics are generated. For higher-order above-bandgap harmonics in the UV range, this depth is shallow due to the small absorption length in the solid, minimizing the role of phase matching. Another macroscopic contribution that has to be considered is the propagation of the driver through the sample, which can also be affected by the control beam. Modulation of the generation beam will subsequently result in harmonic modulation. Both these factors can be controlled by the choice of measurement and sample geometry.

\section{Transient High-Harmonic Generation of Strongly Correlated Materials}

\subsection{Electronic Correlations and the Mott-Hubbard Model}
Thus far, high harmonic generation in solids has been treated mainly within the independent particle approximation, i.e. under the assumption that electrons do not interact with each other: electron-electron interactions are only taken into account in a phenomenological dephasing term, which includes other scattering processes such as scattering off impurities or other collective excitations like phonons or magnons. In general however, electron-electron interactions cannot be neglected in solids. In the Fermi liquid theory, the Coulomb interaction between electrons merely leads to a renormalization of the effective mass, and the electrons are treated as independent particles again.

In cases such as the localized d-orbitals in the transition metals, the Coulomb repulsion between two electrons occupying the same site becomes so strong that it significantly modifies the ground state of the solid and the interaction term needs to be explicitly considered in the many-body Hamiltonian: the simplest approximation is the Mott-Hubbard model \cite{hubbard1963electron}, where the on-site repulsion is taken into account via an energy $U$:
\begin{equation}
\hat H
=
-t \sum_{\langle i,j\rangle,\sigma}
\left(
c_{i\sigma}^\dagger c_{j\sigma}
+
c_{j\sigma}^\dagger c_{i\sigma}
\right)
+
U \sum_i n_{i\uparrow} n_{i\downarrow}.
\end{equation}
When the energy scale of U is much larger than the kinetic energy $t$, the electrons will localize on their respective atomic sites and hopping (conduction) can only occur via the creation of doublon-holon pairs, at the cost of the energy $U$.\\
The strong electron-electron interactions can manifest in different ways in the material properties. 
In Mott insulators, the interaction energy $U$ leads to the formation of a lower and upper Hubbard band, which take the role of valence and conduction bands, respectively. The creation of doublons can then be identified with populating the upper Hubbard band. A simplified energy diagram with doublon excitations indicated is shown in Fig. \ref{fig:HHG_SCMs_static}a. The conductivity of the material depends on the degree of filling of these Hubbard bands: for half filling, i.e. a single electron per unit cell, the lower Hubbard band is fully occupied and the system becomes insulating. Depending on the size of the gap with respect to the ligand energy levels, the material is then classified as either a Mott or a charge-transfer insulator \cite{zaanen1985band}. Such insulating behavior is found in vanadium sesquioxide (\ce{V2O3}) or nickel oxide (\ce{NiO}), among others.\\
In some members of the cuprate family, the strong correlations are believed to be responsible for the high temperature superconducting phase. While at zero doping, these materials are usually Mott insulators, at finite doping concentration, the added carriers can move freely in the copper-oxygen planes and the material even becomes superconducting below the critical temperature.\\
Finally, a strong coupling of the electronic Hamiltonian to the crystal lattice may lead to an insulator-to-metal transition as in \ce{VO2}, \ce{NbO2}, or the rare earth nickelate family \ce{RNiO3}; or a charge density wave (CDW) in some of the transition metal dichalcogenides, for example.\\
\subsection{High-Harmonic Spectra of Mott Insulators}
\begin{figure}[H]
\centering
\includegraphics[width=0.8\linewidth]{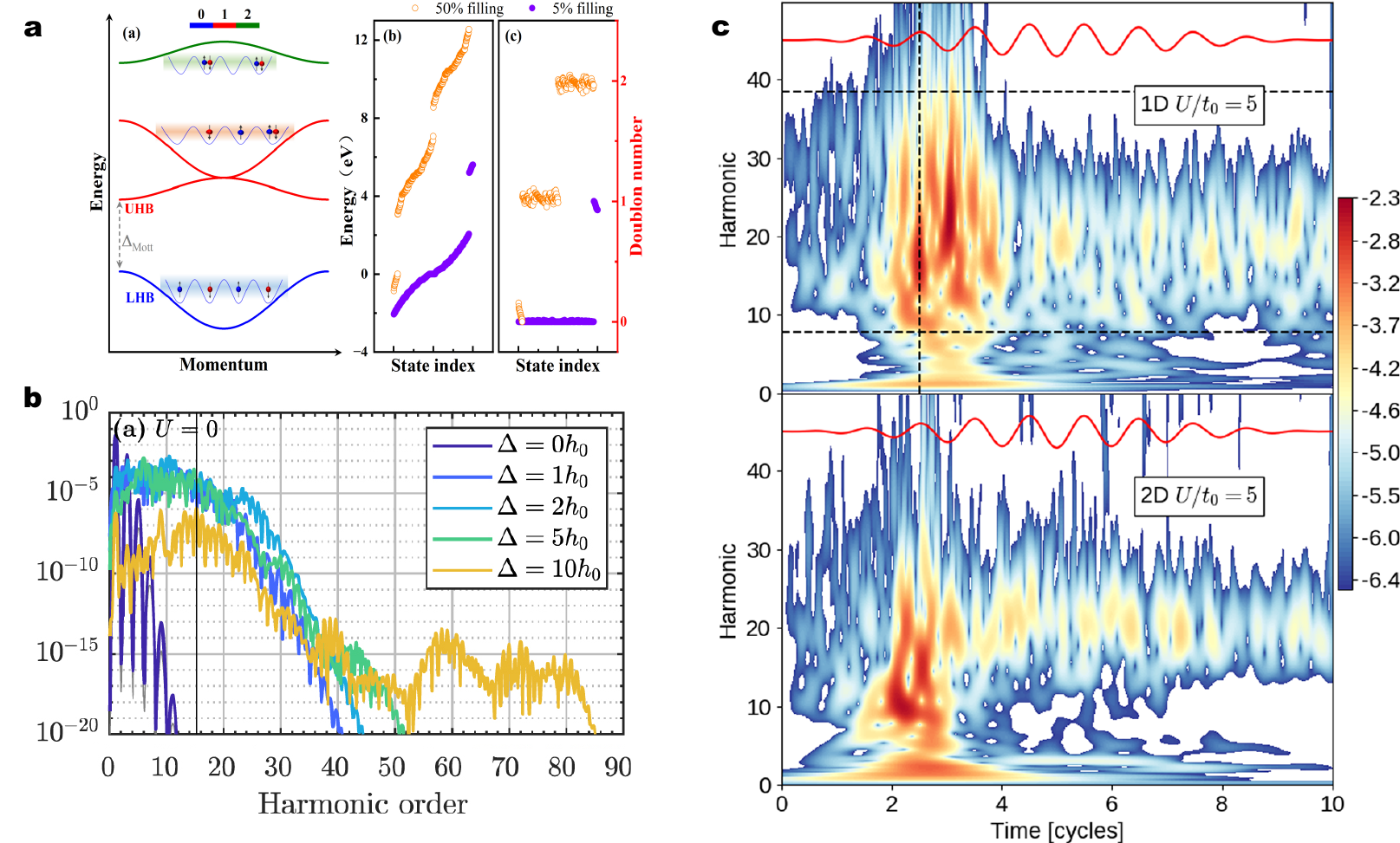}
\caption{(a) Example of the band structure of a Mott insulator, with mapping of doublon/hole states to the conduction and valence bands. (b) Simulated HHG spectra from a Mott insulator, at different values of the interaction constant $U$. (c) Gabor plot of the HHG emission as a function of time over the duration of the generation pulse. The onset of HHG emission corresponds with dielectric breakdown of the Mott insulating state. (a) Reprinted figure with permission from \cite{du2026tracking}. Copyright 2026 by the American Physical Society. (b) Adapted from ref \citenum{hansen2024lattice}. Copyright 2024 IOP Publishing under CC BY 4.0, https://creativecommons.org/licenses/by/4.0/. (c) Adapted from ref \citenum{orthodoxou2021high}. Copyright 2021 Springer Nature under CC BY 4.0, https://creativecommons.org/licenses/by/4.0/.}
\label{fig:HHG_SCMs_static}
\end{figure}

The strong electron-electron correlations are also reflected in the strong-field dynamics and high harmonic emission. First, the different conduction properties of Mott insulators compared to normal band insulators are reflected in their HHG spectra, as will be shown in the following. Secondly, the coupling of electrons to the intense laser pulses needed for HHG can modify the electronic energy landscape on a timescale on the same order as the driving pulse. The resulting ultrafast changes in band structure are then effectively probed by the high harmonic emission, through its sensitivity to the band gap and changes thereof (see equation \ref{eq:dipolephaseapprox}). The ability of exciting ultrafast dynamics combined with the sensitivity to band gap changes therefore makes high harmonic spectroscopy an excellent tool to study the electronic dynamics in strongly correlated materials.\\
The difference in high harmonic spectra between regular and Mott insulators has mainly been investigated theoretically. Several works \cite{Silva2018,orthodoxou2021high,masur2022optical,alshafey2023ultrafast,huang2023quasiparticle} have reported an intensity decrease for the low order harmonic with increasing $U$, while the intensity of high harmonics increases, see Fig. \ref{fig:HHG_SCMs_static}b. The suppression of low-order harmonics is attributed to the restriction of intraband currents in the Mott-insulating state. Instead, the main contribution of HHG comes from doublon-hole recombination, which, similarly to the interband recombination in normal insulators, is responsible for the higher energy harmonics. This interpretation is consistent with the observation of a peak in harmonic emission near the energy scale of the Mott gap $\Delta$, and the onset of high harmonic emission when the driving field is strong enough for nonlinear excitation of doublon-holon pairs \cite{oka2012nonlinear}. Fig. \ref{fig:HHG_SCMs_static}c shows the high harmonic emission after breakdown of the Mott insulating state after approximately two cycles of the driving field.\\
\subsection{Ultrafast Modification of the Interaction Strength and Photoinduced Insulator-to-Metal Transitions}
The interaction strength $U$ is strongly dependent on the orbital configuration of the electrons and the amount of screening they experience from other electrons on the same site. Therefore, $U$ can be modified on a time scale associated with the period of a laser pulse that excites electrons to higher energy states. This was investigated theoretically by \citet{tancogne2018ultrafast} in \ce{NiO} as a model system. It was found that $U$ modulates with twice the frequency of the laser field, in agreement with the hypothesis of reduced screening due to excited electrons. The harmonic spectrum resulting from a dynamic $U$ compared to a frozen $U$ shows clear changes, indicating the sensitivity of HHG to such ultrafast dynamics due to carrier excitations and the resulting band gap shifts.\\
An ultrafast collapse of the band gap can be sensitively probed using HHG spectroscopy due to its sensitivity to the band gap and electron dephasing times. This sensitivity can be exploited in measurements of photo-induced insulator-to-metal transitions (IMT). In Mott or charge-transfer insulators, the photo-excitation of electrons to the conduction band (upper Hubbard band) leads to increased screening of the Coulomb repulsion and subsequent weakening of the insulating state. While an IMT can occur due to this purely electronic change, the influence of the crystal lattice on electronic degrees of freedom can oftentimes not be neglected. A prototypical example of such a combined electronic and structural transition is vanadium dioxide (\ce{VO2}): the weakened electronic interactions destabilize the lattice, and the phase transition is generally considered a combined Mott-Peierls transition \cite{zylbersztejn1975metal}. The timescale of the band gap collapse has been measured to occur in less than 10 fs \cite{brahms2025decoupled}. Upon thermal induction of the phase transition, the intensity of the harmonic emission is observed to decrease, and a similar decrease is also interpreted as a signature of the phase transition in ultrafast pump-probe experiments \cite{Bionta2021, Nie2023,wang2024all}. Although a suppression of harmonic intensity following carrier excitation is generally observed in semiconductor materials, the lasting suppression on picosecond time scales-and increased suppression compared to the electron relaxation dynamics- as shown in Fig. \ref{fig:HHG_SCMs_pumpprobe}a is interpreted as evidence of the photo-induced insulator-to-metal transition.\\
\subsection{Equilibrium Phase Transitions, Spin and Charge Order}
\begin{figure}
\includegraphics[width=0.8\linewidth]{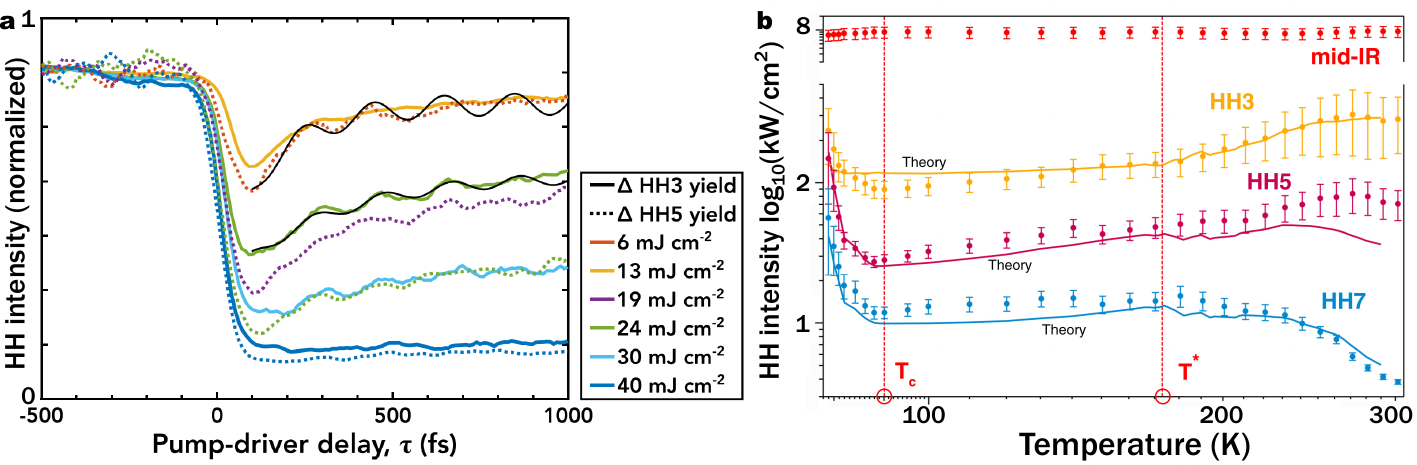}
\caption{(a) Dynamics of the HHG yield from \ce{VO2} for several pump fluences. (b) Temperature-dependent HHG intensity generated in YBCO. (a) Adapted from ref \citenum{Bionta2021}. Copyright 2021 American Physical Society under CC BY 4.0, https://creativecommons.org/licenses/by/4.0/. (b) Adapted from ref \citenum{alcala2022cuprate}. Copyright 2022 National Academy of Sciences under CC BY 4.0, https://creativecommons.org/licenses/by/4.0/.}
\label{fig:HHG_SCMs_pumpprobe}
\end{figure}

Having established high harmonic spectroscopy as a useful tool to study ultrafast dynamics in strongly correlated materials, we now highlight the application of HHG to equilibrium phase transitions in solids.\\
The charge dynamics in strongly correlated materials are strongly coupled to the spin and charge order. When the system is at half filling, an antiferromagnetic ordering of the electron spins reduces the total energy by allowing virtual hoppings between neighboring sites. The interaction of the driven electron with the spin and/or charge background leads to a dephasing that depends on the configuration of the charges or spins. Such an argument was used to explain the temperature dependent high harmonic intensity in Mott-insulating \ce{Ca2RuO4} \cite{uchida2022ruthenate,murakami2022anomalous} and charge-ordered \ce{Pr0.6Ca0.4MnO3} \cite{nakano2024dominant}. As the effective band gap increases with decreasing temperature, the tunneling rate decreases and the high harmonic intensity should be expected to decrease. However, an opposite trend of increased high harmonic intensity with decreasing temperature is measured. The increased high harmonic intensity is attributed to a reduced dephasing due to interactions with the spin background. At low temperature, the increased spin and charge order reduces the dephasing of the quasi-particle trajectories, leading to less destructive interference in the emitted light and therefore a higher intensity. This observation shows how HHG spectroscopy is sensitive to the interactions of electrons with their surroundings on ultrafast timescales.\\
The superconducting phase transition has been studied using HHG spectroscopy in \ce{YBa2Cu3O_{7-\delta}} (YBCO) \cite{alcala2022cuprate}. YBCO features two phase transitions upon cooling from the high-temperature strange metal phase: first to a pseudogap phase and then to the superconducting state. The intensities of harmonics 3 to 7 generated from YBCO are displayed in Fig. \ref{fig:HHG_SCMs_pumpprobe}b and show a clear increase when the material is cooled to the superconducting phase. In addition to the amplitude change, the spectrum exhibits a blueshift with increasing temperature. The blueshift is caused by scattering processes and dephasing during the harmonic generation process. The reconstructed scattering rate is significantly lower in the low temperature phases, reminiscent of the enhanced transport properties.\\
A strong temperature dependence of the harmonic intensity is also observed in CDW system \ce{TiSe2} \cite{tyulnev2025high}. The temperature dependence is anisotropic and depends on the polarization of the laser field with respect to the lattice orientation. The anisotropy below the CDW temperature is linked to an asymmetry of the CDW vectors.


\section{Applications of Controlled and Programmable Solid HHG}

\subsection{Compact and Integrated Short-Wavelength Sources}
The ability of solids to generate high harmonics has naturally motivated strong interest in their use as compact short-wavelength light sources. Compared to gas-phase HHG, solid-state emitters could offer intrinsic advantages in terms of compactness, integrated multi-functional emission targets, and compatibility with nanophotonic and semiconductor fabrication platforms.

At present, however, solid-state HHG sources still remain below state-of-the-art gas-phase systems in terms of achievable cutoff energy, pulse energy, and average XUV power.\cite{allegre25a,Luu2018e,roscam2024} Gas-phase HHG continues to dominate applications requiring isolated attosecond pulses, high photon flux, or access to the soft X-ray regime. Nevertheless, the strength of solid-state HHG lies not primarily in directly competing with large-scale gas-based sources, but rather in enabling functionalities that are difficult or impossible to realize in conventional geometries. In particular, solids provide a route toward integrated and engineerable nonlinear emitters in which generation, shaping, routing, and modulation of short-wavelength radiation can occur within the same structured platform.\cite{Sivis2017,roscam2022} This may be particularly relevant for ultra-compact architectures or applications where lossy XUV optics need to be avoided as much as possible, for instance for squeezed light and pulses with non-classical photon statistics.\cite{rasputnyi2024,lemieux2025}

\begin{figure}[H]
    \centering
    \includegraphics[width=\linewidth]{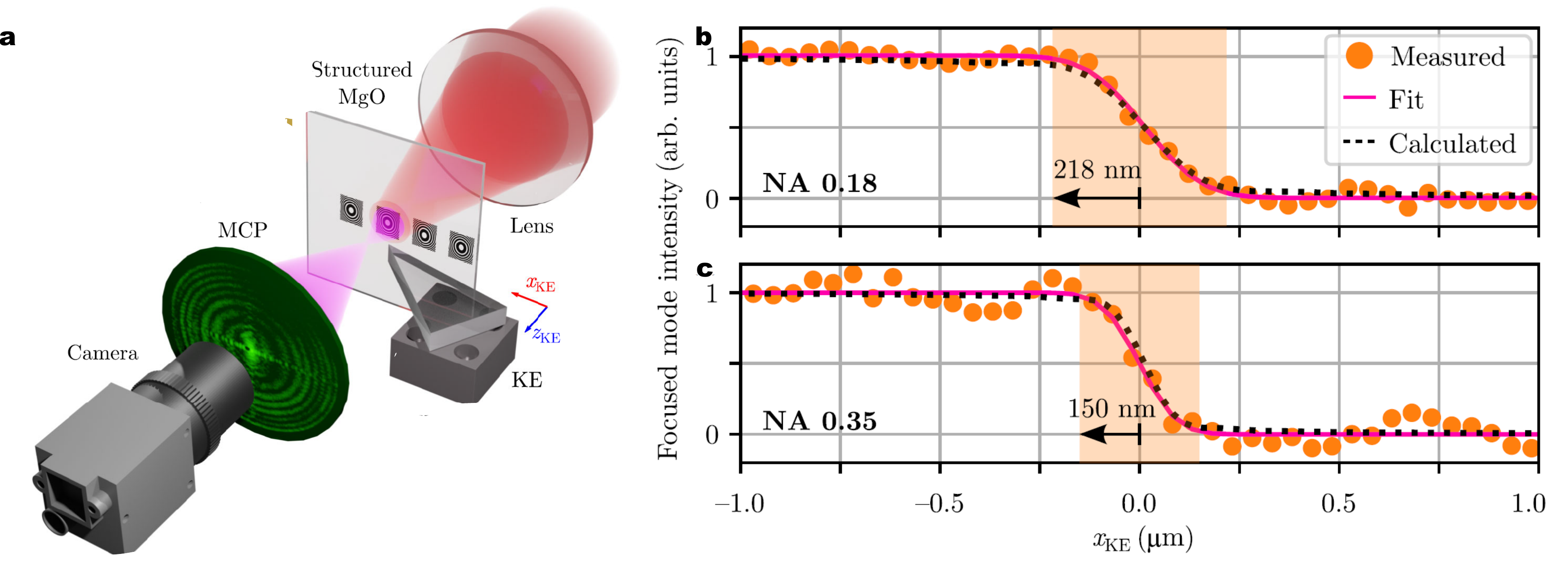}
    \caption{(a) Schematic of the setup of an XUV zone plate engraved in the MgO generation medium, leading to a focusing of the generated light. (b) and (c) show results of the focus measurements for two different numerical apertures, indicated in the figure. Adapted from ref \citenum{Korobenko2022}. Copyright 2022 American Physical Society under CC BY 4.0, https://creativecommons.org/licenses/by/4.0/.}
    \label{fig:XUVZP}
\end{figure}

An important example is the integration of XUV generation and optical functionality into a single solid-state device. Korobenko \emph{et al.} demonstrated that Fresnel zone plates directly etched into an MgO crystal can focus generated XUV harmonics down to nanoscale spot sizes \cite{Korobenko2022}, shown in Fig.~\ref{fig:XUVZP}. In this approach, harmonic generation and focusing occur simultaneously within the same structure, eliminating the need for separate XUV optics. More generally, such demonstrations illustrate how nanostructured solids can merge generation, focusing, shaping, and routing of coherent short-wavelength radiation into unified on-chip platforms.

\subsection{Programmable Harmonic Emission}
Beyond compact source development, the strong dependence of HHG on local electronic structure, excitation conditions, and optical resonances enables the concept of programmable harmonic emission\cite{vanessen2026}. Since harmonic generation is highly sensitive to microscopic excitation pathways, external optical fields can dynamically control where and how harmonics are emitted on femtosecond timescales. In solids, this is particularly powerful because the emitting medium itself can be spatially structured, resonantly enhanced, and locally modified, effectively allowing the nonlinear emission process to become optically reconfigurable.

This ability to locally switch harmonic emission on and off creates interesting parallels to fluorescence-based spectroscopy and microscopy, where optical contrast arises from selectively activated emitters. However, unlike fluorescence, HHG is a coherent process that does not rely on specific molecular transitions, bleaching, or incoherent spontaneous emission. Instead, in principle, HHG can be generated in a very broad range of materials while simultaneously retaining sensitivity to ultrafast electronic dynamics, symmetry, and strong-field interactions. In addition, the emission process itself occurs on femtosecond and sub-cycle timescales, naturally combining spatial selectivity with ultrafast temporal resolution. As a result, programmable HHG may provide a route toward dynamically switchable nonlinear imaging modalities that unite concepts from fluorescence microscopy with the temporal resolution and microscopic sensitivity of strong-field physics.

\begin{figure}[H]
    \centering
    \includegraphics[width=0.6\linewidth]{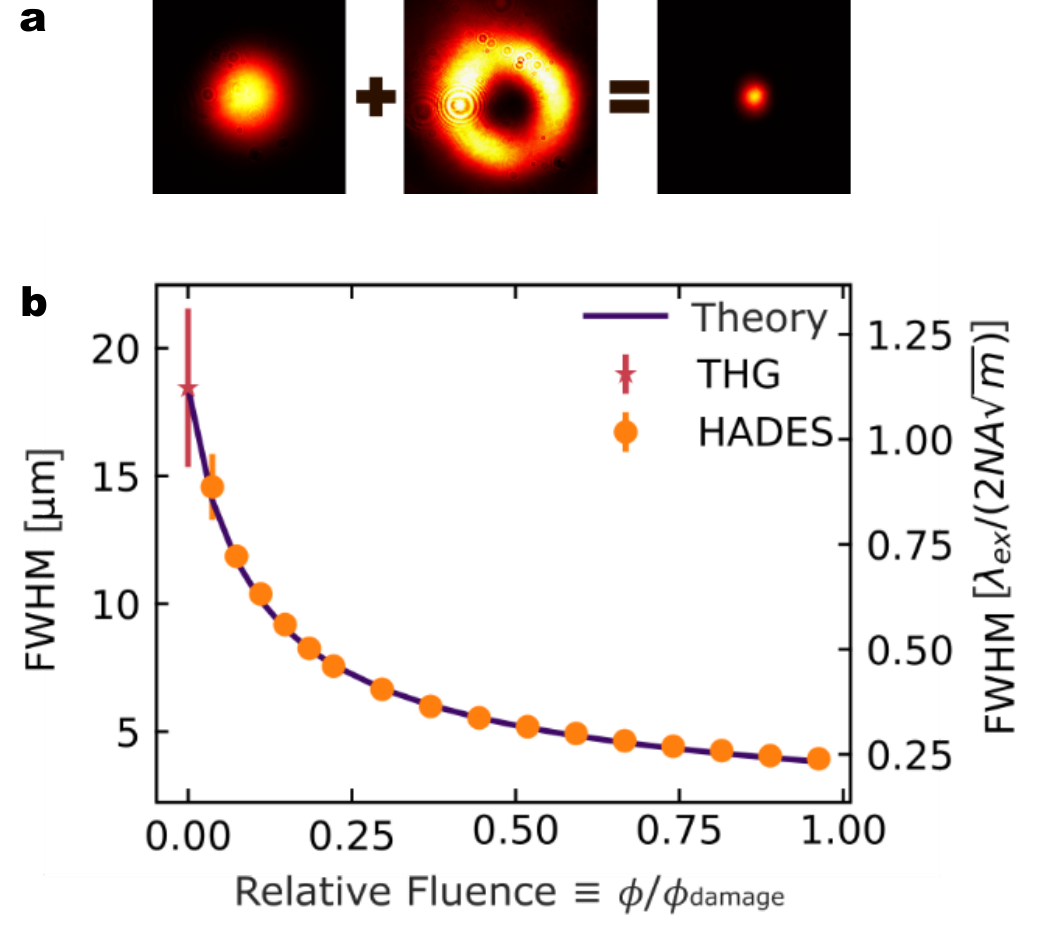}
    \caption{(a) Microscopy image of (left) the third harmonic from NbO$_2$ at 667~nm, (center) the donut-shaped pre-pulse centered at 800~nm and (right) the resulting harmonic emission obtained by combining both pulses. (b) Scaling of the width of the third harmonic emission in HADES for increasing fluence of the donut pre-pulse. In all panels, the beams were loosely focused to allow resolving with a wide-field microscope, and the NA in (b) refers to the NA of the loose focusing lens. Adapted from refs \citenum{Murzyn2024} and \citenum{murzyn2026}. Copyright 2024 American Association for the Advancement of Science and Copyright 2026 Optica Publishing Group under CC BY 4.0, https://creativecommons.org/licenses/by/4.0/.}
    \label{fig:HADES}
\end{figure}

\subsection{Super-Resolution Microscopy with Switchable Harmonics}
This concept directly motivates HHG-based imaging and microscopy schemes such as HADES\cite{Murzyn2024,murzyn2026}, where harmonic emission itself acts as the imaging observable. Similar to fluorescence microscopy, contrast is not generated by transmitted probe light alone, but rather through localized nonlinear emission from the sample. Sub-diffraction localization is achieved by deactivating HHG with a donut-shaped pre-pulse, which confines HHG to its center where it carries no intensity, as demonstrated in Fig.~\ref{fig:HADES}. While ultimately limited by sample damage, this concept recovers a resolution scaling reminiscent of stimulated-emission-depletion (STED) microscopy, where the achievable resolution improvement scales inversely with the square root of the fluence applied in the donut prepulse relative to the saturation fluence, which is the fluence where half the signal is deactivated. Thus, the resolution improvement in HADES microscopy follows directly from the observables of time-resolved high-harmonic spectroscopy. The programmability to completely switch off HHG with another laser pulse thus takes a similar significance as switching off fluorescence via stimulated emission. Thus, many other spectroscopy and microscopy techniques that previously relied on this fluorescence property could be reinvented via the programmability of HHG from solids.

\subsection{Dielectric Metasurfaces as Programmable Nonlinear Emitters}
While HHG in solids is generally switchable \cite{vanessen2026,essen2025spatial}, another platform for such concepts are structured media and, in particular, dielectric metasurfaces. Nanostructuring the generation medium itself was shown early on to enhance and shape harmonic emission from semiconductors and dielectrics \cite{Sivis2017,Franz2019a,roscam2022}, and resonant metasurfaces have since become a platform in their own right. Enhanced harmonic generation was demonstrated from an all-dielectric metasurface \cite{Liu2018a}, even and odd harmonics were generated in resonant metasurfaces driven by single and multiple intense pulses \cite{Shcherbakov2021a}, and resonances engineered from bound states in the continuum were used to boost the conversion efficiency further \cite{Zograf2022}. Metasurfaces can therefore strongly enhance nonlinear conversion efficiencies through local field enhancement while simultaneously imprinting spatial and spectral functionality onto the emitted harmonics \cite{bijloo2024,bijloo2026}. Because the harmonic emission is generated coherently from structured nanoscale emitters, the emitted wavefront can be directly engineered at the generation stage, enabling control over diffraction, directionality, polarization, and Fourier-space emission.

Recent experiments demonstrated a nearly unity all-optical modulation of harmonic generation in Fano-resonant dielectric metasurfaces through ultrafast carrier-induced resonance detuning \cite{bijloo2024}. Building on this concept, dynamically structured pump fields were subsequently used to spatially program harmonic emission in both real and Fourier space \cite{bijloo2026}. Spatially patterned excitation locally deactivated harmonic generation across the metasurface, allowing arbitrary images, diffraction patterns, and directional emission states to be imprinted directly onto the emitted harmonics. Importantly, because the HHG process remains coherent, this modulation affects both near-field and far-field emission properties. These results illustrate how metasurfaces can evolve from static nonlinear emitters toward dynamically programmable nonlinear optical platforms.

More broadly, these developments illustrate a new avenue for research in the field. Initially, solid-state HHG was primarily viewed as a spectroscopic observable encoding electronic structure and ultrafast dynamics. However, in a growing number of cases, HHG is evolving into an actively engineered nonlinear process, where microscopic excitation pathways, interference effects, and optical resonances are deliberately manipulated to realize tailored short-wavelength emission with designed spatial, spectral, and temporal properties.

\section{Conclusion and Outlook}

High-harmonic generation in solids has evolved from an initially unexpected strong-field phenomenon into a versatile platform for studying, controlling, and engineering ultrafast electronic dynamics. As discussed throughout this review, HHG simultaneously provides access to band structure, coherence, scattering, symmetry, topology, and many-body interactions, while its extreme sensitivity to excitation conditions enables direct optical control of the emission process itself. This dual role, as both spectroscopic observable and controllable nonlinear process, increasingly distinguishes solid-state HHG from its gas-phase counterpart.

At the same time, many fundamental questions remain open. In particular, the microscopic origin of dephasing, the interplay between electronic correlations and strong-field dynamics, and the role of propagation and collective effects still require substantially deeper understanding. Progress will likely depend on combining multidimensional experiments with increasingly realistic microscopic modeling that bridges strong-field physics, condensed matter theory, and nonlinear optics.

From an applications perspective, solid-state HHG may ultimately find its greatest impact by enabling functionalities unique to structured and integrated materials platforms. The emerging ability to program, switch, and spatially engineer harmonic emission suggests future directions toward ultracompact XUV photonics, adaptive nonlinear optics, ultrafast imaging via super-resolution concepts, and petahertz optoelectronics. In this sense, the field is exploring strategies to actively design nonlinear emission processes themselves.





\bibliography{BIB_ACSbook_revised}

\end{document}